
\input harvmac
\def\a{\alpha}    \def\b{\beta}              \def\d{\delta}
\def\D{\Delta}    \def\e{\varepsilon}        
\def\g{\gamma}    \def\G{\Gamma}           \def\l{\lambda}
\def\L{\Lambda}   \def\m{\mu}         \def\n{\nu}        
\def\vr{\varrho}        \def\O{\Omega}     \def\p{\psi}
\def\P{\Psi}      \def\s{\sigma}      \def\S{\Sigma}     
        \def\w{\varphi}    

\def\CA{{\cal A}}

\def\CG{{\cal G}}
\def\CM{{\cal M}}
\def\CN{{\cal N}}

\def\CL{{\cal L}}
\def\CD{{\cal D}}

 \def\CR{{\cal R}}
\def\CT{{\cal T}}
%
\font\teneufm=eufm10
\font\seveneufm=eufm7
\font\fiveeufm=eufm5
\newfam\eufmfam
\textfont\eufmfam=\teneufm
\scriptfont\eufmfam=\seveneufm
\scriptscriptfont\eufmfam=\fiveeufm
\def\eufm#1{{\fam\eufmfam\relax#1}}

\font\teneusm=eusm10
\font\seveneusm=eusm7
\font\fiveeusm=eusm5
\newfam\eusmfam
\textfont\eusmfam=\teneusm
\scriptfont\eusmfam=\seveneusm
\scriptscriptfont\eusmfam=\fiveeusm

\font\tenmsx=msam10
\font\sevenmsx=msam7
\font\fivemsx=msam5
\font\tenmsy=msbm10
\font\sevenmsy=msbm7
\font\fivemsy=msbm5
\newfam\msafam
\newfam\msbfam
\textfont\msafam=\tenmsx  \scriptfont\msafam=\sevenmsx
  \scriptscriptfont\msafam=\fivemsx
\textfont\msbfam=\tenmsy  \scriptfont\msbfam=\sevenmsy
  \scriptscriptfont\msbfam=\fivemsy

\def\msbm#1{{\fam\msbfam\relax#1}}


\def\rd{\partial}

\def\darr#1{\raise1.5ex\hbox{$\leftrightarrow$}\mkern-16.5mu #1}

\def\Fr#1#2{{#1\over#2}}
\def\tr{\hbox{Tr}\,}

\def\Fs#1{#1\!\!\!\!/\,} 
\def\roughly#1{\raise.3ex\hbox{$#1$\kern-.75em\lower1ex\hbox{$\sim$}}}

%
\def\cmp#1#2#3{Comm.\ Math.\ Phys.\ {{\bf #1}} {(#2)} {#3}}
\def\pl#1#2#3{Phys.\ Lett.\ {{\bf #1}} {(#2)} {#3}}
\def\np#1#2#3{Nucl.\ Phys.\ {{\bf #1}} {(#2)} {#3}}

\def\ijmp#1#2#3{Int.\ J.\ Mod.\ Phys.\ {{\bf #1}} {(#2)} {#3}}

\def\top#1#2#3{Topology {{\bf #1}} {(#2)} {#3}}

\def\prp#1#2#3{Phys.\ Rep.\ {{\bf #1}} {(#2)} {#3}}

\def\plms#1#2#3{Proc.\ London Math.\ Soc.\ {{\bf #1}} {(#2)} {#3}}

\def\bams#1#2#3{Bull.\ Am.\ Math.\ Soc.\ {{\bf #1}} {(#2)} {#3}}
\def\jgp#1#2#3{J.\ Geom.\ Phys.\ {{\bf #1}} {(#2)} {#3}}

\def\pr{\prime}

\def\gBE{\eufm{g}_{\raise-.1ex\hbox{${}_\BE$}}}
\def\gEc{\eufm{g}_{\raise-.1ex\hbox{${}_E$}}^C}

\def\BE{\msbm{E}}

\def\BR{\msbm{R}}
\def\BZ{\msbm{Z}}

\lref\Donaldson{
S.~Donaldson,
{\it Polynomial invariants for smooth four-manifolds},
\top{29}{1990}{257}.
}
\lref\DK{
S.K.\ Donaldson and P.B.\ Kronheimer,
{\it The geometry of four-manifolds},
(Oxford University Press, New York, 1990)
}
\lref\REVa{
D.\ Birmingham, M.\ Blau, M.\ Rakowski and G.\ Thomson,
{\it Topological field theory},
\prp{209}{1991}{129}
}
\lref\REVb{
S.~Cordes, G.~Moore and S.~Ramgoolam,
{\it Lectures on 2D Yang-Mills theory, equivariant cohomology
and topological field theories}, Part II,  hep-th/9411210.
}
\lref\WittenA{
E.~Witten,
{\it Topological quantum field theory},
\cmp{117}{1988}{353};
{\it Introduction to cohomological field theories},
\ijmp {A 6}{1991}{2273}.
}
\lref\WittenB{
E.~Witten,
{\it Supersymmetric Yang-Mills theory on a four manifolds},
J.~Math.~Phys.~{\bf 35} (1994) 5101.
}
\lref\WittenC{
E.~Witten,
{\it Monopoles and four-manifolds},
Math.~Research Lett.~{\bf 1} (1994) 769.
}
\lref\WittenD{
E.~Witten,
{\it The $N$ matrix model and gauged WZW models},
\np {B 371}{1992}{191}.
}
\lref\WittenE{
E.~Witten,
{\it Two dimensional gauge theories revisited},
\jgp{9}{1992}{303}.
}
\lref\WittenS{
E.~Witten,
{\it On $S$-duality in abelian gauge theory},
hep-th/9505186.
}
\lref\SWa{
N.~Seiberg and E.~Witten,
{\it Electric-magnetic duality, monopole condensation, and confinement
in $N=2$ supersymmetric Yang-Mills theory},
\np{B 426}{1994}{19}.
}
\lref\SWb{
N.~Seiberg and E.~Witten,
{\it Monopoles, duality, and chiral symmetry breaking in
$N=2$ supersymmetric QCD},
\np{B 431}{1994}{484}.
}
\lref\KMa{
P.~Kronheimer and T.~Mrowka,
{\it Recurrence relations and asymptotics
for four-manifold invariants},
\bams{30}{1994}{215}.
}
\lref\HPPa{
S.~Hyun, J.~Park and J.-S.~Park,
{\it Topological QCD}, Nucl.~Phys.~B to appear,
hep-th/9503201.
}
\lref\HPPb{S.~Hyun, J.~Park and J.-S.~Park,
{\it The $N=2$ Supersymmetric QCD
and Four Manifolds; (II) the $SU(N_c)$ and $N_f =0, 1$ cases,
} to appear.
}
\lref\ML{
B.~Lawson, JR and M-L. Michelsohn,
{\it Spin geometry}, (Princeton University Press, 1989).
}
\lref\ANF{
D.~Anselmi and P.~Fr\'{e},
{\it Topological twist in four dimensions, R-duality and hyperinstantons},
\np{B 404}{1993}{288}; {\it Topological $\s$-models in
four dimensions and triholomorphic maps}, \np{B 416}{1994}{255};
{\it Gauged hyperinstantons and monopole equations},
hep-th/9411205.
}
\lref\BGV{
N.~Berline, E.~Getzler and M.~Vergne,
{\it Heat kernels and Dirac Operators},
(Springer-Verlag, 1992).
}
\lref\AJ{
M.F.~Atiyah and L.~Jeffrey,
{\it Topological Lagrangians and cohomology},
\jgp{7}{1990}{1}.
}
\lref\AB{
M.F.~Atiyah and R.~Bott,
{\it The moment map and equivariant cohomology},
\top{23}{1984}{1}.
}
\lref\DH{
J.J.\ Duistermaat and G.J.\ Heckmann,
{\it On the variation in the cohomology
in the symplectic form of the reduced phase space},
Invent Math.\ {\bf 69} (1982) 259;
{\it Addendum},
Invent Math.~{\bf 72} (1983) 153.
}
\lref\LM{
J.M.F.~Labastida and M.~Mari$\tilde{\rm n}$o,
{\it Non-abelian monopoles on four-manifolds}
hep-th/9504010
}
\lref\Lc{
M.~Alvarez and J.M.F.~Labastida,
{\it Breaking of topological symmetry},
\pl{B 315}{1993}{251}.
}
\lref\Ld{
M.~Alvarez and J.M.F ~Labastida,
{\it Topological matter in four dimensions},
\np{B 437}{1995}{356}.
}
\lref\Hitchin{
N.J.~Hitchin,
{\it The self-duality equations on a Riemann Surface},
\plms{55}{1987}{59}.
}
\lref\VW{
C.~Vafa and E.~Witten,
{\it A strong coupling test of $S$-duality},
\np{B 431}{1994}{3}.
}
\lref\KMT{
P.~Kronheimer and T.~Mrowka,
{\it The genus of embedded surfaces in the
projective plane},
preprint 1994.
}
\lref\Taubes{
C.~Taubes,
{\it Symplectic manifolds and the Seiberg-Witten invariants},
preprint, 1994.
}
\lref\PT{V.~Pidstrigach and A.~Tyurin,
{\it Lectures given at the Isaac Newton Institute}, Dec.~1994}
\lref\FS{
R.~Fintushel and R.J.~Stern,
{\it The blowup formula for Donaldson invariants},
alg-geom/9405002.
}
\lref\PTb{
V.~Pidstrigach and A.~Tyurin,
{\it Localisation of the Donaldson's invariants along
Seiberg-Witten classes},
dg-ga/9507004
}
\lref\LMb{
J.M.F.~Labastida and M.~Mari$\tilde{\rm n}$o,
{\it Polynomial invariants for $SU(2)$ monopoles},
hep-th/9507140
}

\baselineskip=15pt plus 1.2pt minus .6pt
\newskip\normalparskip
\normalparskip = 8pt plus 1.2pt minus .6pt
\parskip = \normalparskip
\parindent=18pt

\def\ack{\bigbreak\bigskip\bigskip\centerline{{\bf Acknowledgements}}\nobreak}
\font\Titlerm=cmr10 scaled\magstep2
\nopagenumbers
\rightline{YUMS-95-13, CALT-68-1995,  SWAT/69}
\rightline{hep-th/9508162}
\vskip .4in
\centerline{\fam0\Titlerm N=2 Supersymmetric QCD and Four Manifolds;}
\vskip .1in
\centerline{\bf (I)  the Donaldson and the Seiberg-Witten Invariants}
\tenpoint\vskip .4in\pageno=0

\centerline{
Seungjoon Hyun
}
\centerline{{\it Institute for Mathematical Sciences, Yonsei University}}
\centerline{{\it Seoul 120-749, Korea}}
\centerline{{\it (hyun@phya.yonsei.ac.kr)}}
\centerline{and}
\centerline{
Jaemo Park
}
\centerline{\it Department of Physics, California Institute of Technology}
\centerline{\it Pasadena, CA 91125, USA}
\centerline{\it (jaemo@cco.caltech.edu)}
\centerline{and}
\centerline{
Jae-Suk Park
\footnote{$^{\dagger}$}{
Address after 1 October 1995 : Institute for Theoretical Physics,
University of Amsterdam, Valckenierstraat 65,  1018 XE Amsterdam,
The Netherlands}
}
\centerline{\it Department of Physics, University of Wales, Swansea}
\centerline{\it Swansea SA2 8PP, UK }
\centerline{\it (j.park@swansea.ac.uk)}
\vskip .3in

\noindent
\abstractfont
We study the path integral of a twisted $N=2$ supersymmetric
Yang-Mills theory coupled with hypermultiplet having
the bare mass.  We explicitly compute the topological correlation
functions  for the $SU(2)$ theory on a compact oriented
simply connected   simple type Riemann manifold with $b_2^+ \geq 3$.
As the corollaries, we determine
the topological correlation functions of the theory
without the bare mass and those of the theory
without coupling to the hypermultiplet.
This includes a concrete field theoretic proof of
the relation between the Donaldson and the Seiberg-Witten
invariants.
\tenpoint

\Date{July, 1995}
\baselineskip=14pt plus 1.2pt minus .6pt

\newsec{Introduction}

Both the Donaldson and the Seiberg-Witten invariants of
smooth four-manifolds  are closely related to the  $N=2$
supersymmetric Yang-Mills (SYM) theory \WittenC.

The cohomological description of the Donaldson theory \Donaldson\
(for review, see \DK)
is represented by the twisted version of  $N=2$ SYM
theory \WittenA, called the topological Yang-Mills (TYM)
theory  (for reviews and
references, see \REVa\REVb).  The semi-classical
analysis of the TYM theory,
based  on the ultraviolet weak coupling limit
of the underlying physical theory,  was used to
reformulate the Donaldson theory in a
concrete  way.  Those cohomological descriptions, however,
turned out to be surprisingly difficult to obtain explicit results
{}.

On the other hand, the physical interpretation of Witten
opened the door to an entirely different formulation of
the Donaldson theory. This is due to the asymptotic freedom
of the underlying physical theory.  In the infrared or the
large scaling limit, the physical theory is strongly coupled
and the semi-classical description is not valid.  Since the
TYM theory is metric independent, the Donaldson theory
can be reformulated in terms of the new degrees of freedom
that may appear in the strong coupling vacua.
In a seminal paper \SWa,  Seiberg and Witten determined
the exact infrared behavior of the  $N=2$ SYM theory.
The celebrated Seiberg-Witten invariant
originates from the resulting low-energy
effective theory \WittenC, which is  a simple and
powerful new tool for the study of differential-topology of
four-manifold \KMT\Taubes.

For an oriented simply connected compact Riemann four-manifold $X$
of simple type  \KMa\ with $b_2^+ \geq 3$,
Witten conjectured a precise formula
relating the $SU(2)$ Donaldson invariants  with the Seiberg-Witten
invariants\WittenC
\eqn\goal{\eqalign{
\left<\exp\left( \hat v + \tau \hat u\right)\right>
=
& 2^{1 + \Fr{1}{4}(7\chi + 11\s)}
   \biggl(\exp\left(\Fr{v\cdot v}{2} + 2\tau \right)\sum_x n_x e^{v\cdot x}
\cr
& \phantom{........}+ i^{(\chi +\s)/4}\exp
   \left(-\Fr{v\cdot v}{2} - 2\tau \right)\sum_x n_x e^{-iv\cdot x}
   \biggr),
\cr
}}
where $x$ is the Seiberg-Witten basic class,
$n_x$ is the algebraic sum of the number of the solutions
of the Seiberg-Witten equation and $v\in H_2(X;\BZ)$.
The above formula agrees with the structure formula
of Kronheimer-Mrowka for the simple type manifold
as well as with the results of  the paper \WittenB\
where the Donaldson invariants on K\"{a}hler surface
was determined almost completely using the known
vacuum structure of $N=1$ SYM theory.
Some progress in a mathematical proof
has been announced  by  Pidstrigach and Tyurin \PT.
More recently, Witten explained the appearance of the
$spin^c$ in the low energy effective theory \WittenS.

In our previous paper \HPPa, we showed that
the $N=2$ SYM theory
coupled with hypermultiplet (SQCD) can be twisted
after picking a $spin^c$ structure $\eufm{c}$
to define a global supersymmetric theory
called  topological QCD (TQCD) on an arbitrary
oriented Riemann four-manifold $X$.
TQCD is  a  generalization  of  the TYM theory.
TQCD shares many of the properties of the TYM theory
as a cohomological field theory in which a suitable
path integral defines differential-topological invariants of
smooth four-manifold.
The topological amplitudes of  TQCD are the
intersection pairings,
analogous to the Donaldson invariants,
in the moduli space $\CM(k,\eufm{c})$ of the non-abelian version of
the Seiberg-Witten monopoles.

We note that  the twisting of the general  $N=2$ hypermultiplets
including gravity was previously  studied  by Anselmi and Fr\'{e}
using the  $\s$-model interpretation \ANF.
On $K3$ surface, their theory is equivalent to TQCD
with the choice of trivial $spin^c$ strucuture.
Based on the  Mathai-Quillen formalism,
Labastida and Mari\~{n}o constructed a topological  theory
on spin manifold \LM\  which is equivalent to  TQCD  with
the choice
of the trivial $spin^c$  structure.

At first sight, TQCD looks like just a clone of
the TYM theory after replacing the moduli space $\CM(k)$ of
anti-self-dual (ASD) connections with the moduli space
$\CM(k,\eufm{c})$.
The cohomological interpretation of the invariants is also based
on the weak coupling  limit of the underlying physical theory.
The theory also
shares the notorious difficulty for explicit computations
with the TYM theory. Furthermore,   those invariants defined
by TQCD  appear not to contain any information  beyond the
Donaldson and the Seiberg-Witten invariants.

However,  TQCD has an interesting property.
By introducing the $N=2$ supersymmetric bare mass term
to the hypermultiplet, we will show that
the resulting TQCD  (massive TQCD) interpolates
the Donaldson and the Seiberg-Witten theories.
Our picture rather contrasts   with
the approach of  Seiberg and Witten in that
the genuine quantum scaling behavior of the underlying
$N=2$ SYM   theory interpolates  the above two limits \SWa.
On the other hand, we consider a weak coupling limit
of a different asymptotically free theory after turning on the bare mass
for the additional matter-multiplet.  Our computation does
not use any known structures of the vacua or the electro-magnetic
duality of the underlying physical $N=2$ supersymmetric theories.

In this paper, we study the $SU(2)$ theory with the hypermultiplet
carrying the fundamental representation. We determine
the topological correlation functions of massive TQCD on a simple
type manifold. As the corollaries, we derive the topological correlation
functions of the theory without the bare mass as well as those
of the theory without hypermultiplet (TYM theory).
The last result is a simple and perfectly concrete
path integral  proof of the formula \goal.

Introducing the bare mass term to the hypermultiplet,
we show that the path integral is localized to
two different branches;
in branch (i) the dominant contribution comes from the
moduli space of ASD connections, while in  branch (ii) the
dominant contribution comes from the moduli space of
the (abelian) Seiberg-Witten monopoles.
To put it differently,  branch (i)
is governed by the TYM theory and  branch (ii) is governed by
a topological QED coupled with massless hypermultiplet.
We show the above property by using Witten's fixed point
theorem for the global supersymmetry \WittenD\ as well as
by the semi-classical analysis combined with
an additional stationary phase method
after  adding a BRST trivial deformation related
to the bare  mass term. It turns out that a key simplication
occurs in  the large scale limit of the metric.
Our computation is rather elementary using the simple
Gaussian integrals. In the subsequent paper, we will give the
the explicit computations of
topological invariants defined  by $SU(N_c)$
TYM and TQCD with $N_f = 1 $ hypermultiplets carrying
the fundamental representation and clarify their relations
with the underlying physical theories \HPPb.

This paper is organized as follows.
In  section 2, we briefly review  TQCD without the bare mass \HPPa.
The purpose of this section is
to establish our notations and to make this paper reasonably
self-contained. We  add some important  remarks
as well. In  section 3,
we study the twisted theory after introducing
the bare mass term to the hypermultiplet.
Adopting various arguments we show that the path integral
of the resulting theory is localized to two types of the branches.
Then, we show that the key step is to take the large scale
limit of the Riemann metric.
In  section 4, the path integrals of TQCD
having the bare mass is  computed
in the large scale limit. As the corollaries, we prove the formula
\goal\ and obtain the precise formula for the invariants defined by
TQCD with an arbitrary $spin^c$ structure.
In section 5, we briefly discuss some relations
with the physical theory.

The appendix is devoted to a demonstration of
the path integral method which is a slightly simpler
version of the technique we used in the actual
computations.

\newsec{The  Topological QCD}

In this section, we briefly review
the $N=2$ global (rigid) supersymmetric
Yang-Mills theory coupled with hypermultiplet
(TQCD) on the general oriented
compact Riemann $4$-manifolds \HPPa.

\subsec{Twisting}

In the  flat $4$-manifold $\BR^4$, the $N=2$ rigid supersymmetric
theories are well-defined. The theory contains an $N=2$
vector multiplet in the adjoint representation.
In addition,  one can couple
$N=2$ matter multiplets  known as the hypermultiplets
carrying a representation $R$ and its conjugate representation
$\tilde R$ of the gauge group.
Those multiplets contain various spinor fields
which are well defined.

Now we consider a compact oriented simply connected Riemann four
manifold $X$.  We want to define the supersymmetric theory on $X$
without destroying the global (rigid) supersymmetry or,
equivalently, without introducing the dynamical gravity.
Then there are two obstructions;

(a) The non-existence of the spinor fields on the manifold $X$ with
    the non-vanishing second Stiefel-Whitney $w_2(X)\neq 0$
    class.

(b) The non-existence of the nowhere vanishing constant supersymmetry
     charge.

The obstruction (b) is generic since the supercharges
transform as the spinor  and the rigid supersymmetry requires
the existence of  nowhere vanishing  constant spinor.
Thus, the rigid supersymmetric theory
can be defined only on the parallelizable
(topologically trivial) manifolds.
On a topologically non-trivial manifold,
the only quantity that can be defined
as a nowhere vanishing constant is the one which transforms
as the scalar.   The twisting procedure introduced by Witten
is the recipe to make the supercharge transform
as the scalar \WittenA\WittenB.

For the $N=2$ vector multiplet, the twisting  resolves
obstructions (a) and (b) simultaneously.
After twisting,  there are no fields which transform as the spinor.
And,  the supercharges have components which transform as the vector,
the self-dual tensor and the scalar.
If we take the component which transforms as the scalar
and use it as the new global supercharge,
the obstruction (b) is removed as well.
One important remark is that the resulting
action functional is independent of
the spin structure on $X$ and  the Riemann curvature
of the manifold.

Twisting of the hypermultiplet is slightly more subtle.
Since the twisted supercharge transforms as a scalar, the obstruction
(b)  does not exist.
After twisting,
every component field of hypermultiplet transforms as spinor.
If the manifold has spin structure, i.e., $w_2(X)=0$,
the resulting theory will be well-defined.
Unlike the theory without hypermultiplet, the twisted theory depends
on the spin structure.\foot{A simply connected spin manifold
has  unique spin structure.}
If the manifold $X$ does not admit
a spin structure the twisted theory may not exist.
A practical way of resolving the obstruction (a) is to
regard the twisted hypermultiplet as $spin^c$ spinor
which exists in any oriented Riemann $4$-manifold.
Roughly speaking, the $spin^c$ spinor
transforms as the spinor with certain background $U(1)$ charge
whose connection couples with the Dirac operator.
The additional $U(1)$ connection compensates the
obstruction to define the spinor.  A $spin^c$ structure
$\eufm{c} \in H^2(X;\BZ)$
is an integral lift of the Stifel-Whitney class
$w_2(X) \in H^2(X;\BZ/2)$, i.e.
$\eufm{c} \equiv w_2(X) \hbox{ mod } 2$ \ML.
The price we should pay is that the twisted theory
depends on a particular $spin^c$ structure
we choose.
The space $H^2(X;\BR)$ of
harmonic two-forms on $X$ is an
$b_2$-dimensional flat  space
with signature  $(b_2^+ - b_2^-)$. The space $H^2(X;\BZ)$ is
the integral lattice in $H^2(X;\BR)$. Then, the set $H^2_s(X;\BZ)$
of all $spin^c$ structure is an affine sublattice of $H^2(X;\BZ)$.
Obviously, there are $b_2$ independent generators $\eufm{s}$ of
the transitive action on $H^2_s(X;\BZ)$.
Thus we have a family of the TQCD
parametrized by the space $H^2_s(X;\BZ)$ of the $spin^c$ structures
on $X$. It would be worthy to remark that even in
a spin manifold one can consider the theory
defined by an arbitrary $spin^c$ structure.
In our viewpoint, the only difference between the spin and non-spin
manifold is that the spin manifold has the $spin^c$
bundle with the trivial  determinant line bundle
(trivial background $U(1)$ connections).

Let $P$ be a principal $G$-bundle over a simply connected
compact oriented Riemann manifold $X$.
Pick a representation $R$ of $G$ such that
$c_2(adj) -T(R) \geq 0$ and consider the
associated vector bundle $E$ whose fiber
is the vector space $V$, $R: G\rightarrow V$.
We pick a $spin^c$ structure
$\eufm{c}$ on $X$ and consider the associated $spin^c$ bundle
$W^\pm_\eufm{c}$.\foot{
We will occasionally confuse with the
determinant line bundle $det(W^\pm_\eufm{c}) =L_\eufm{c}$
and its first Chern class $c_1(det(W^\pm_\eufm{c}))$.
Both will be denoted by $\eufm{c}$.
To avoid possible confusion we will always denote
the intersection number explicitly using $'\cdot'$.
For example, $\eufm{c}\cdot\eufm{c} = L_\eufm{c}\cdot L_\eufm{c}
= c_1(L_\eufm{c})\cdot c_1(L_\eufm{c}) =
\int_X c_1(L_\eufm{c})\wedge c_1(L_\eufm{c})$.
The expression $\eufm{c}^2$ can mean $L_\eufm{c}^2$ or  $2\eufm{c}$.
Note also that
$\eufm{c} + 2\zeta \equiv L_\eufm{c}\otimes L_\zeta^2$.
}
Let $\CA$ be the space of all connections on $P$ and
$\G(W^+_\eufm{c}\otimes E)$  the space of the sections of the $spin^c$
bundle twisted by the vector bundle $E$.
After twisting, the  complex boson (squark) in the hypermultiplet
become a section of  $W^+_\eufm{c}\otimes E$;
\eqn\yaa{
q \in \G(W^+_\eufm{c}\otimes E),\qquad
q^\dagger \in \G(\overline{W}^+_\eufm{c}\otimes \tilde{E}),
}
where $\tilde{E}$ denotes the vector bundle conjugate to $E$.
The $spin^c$ Dirac operator
\eqn\yab{
\s^\m D_\m :\G(W^+_\eufm{c}\otimes E)\rightarrow
\G(W^-_\eufm{c}\otimes E),
}
is the Dirac operator for the $spin^c$ bundle twisted by $E$.
We will sometimes denote $\s^\m D_\m$ by  $\Fs{D}$
or by $\Fs{D}_{\eufm{c}}^{E}$.
One effect of twisting the hypermultiplet is that
the Dirac operator is coupled with the background $U(1)$
connection of the determinant line bundle $det(W_\eufm{c}^+)$.
This can be summarized by the Weitzenb\"{o}ck formula;
\eqn\wzbf{
(\Fs{D})^2 = -g^{\m\n}D_\m D_\n - F^+_A - p^+ +\Fr{1}{4}R,
}
where $F^+_A$ is the self-dual part of the gauge field strength,
$p^+$ denotes the self-dual part of the curvature $2$-form
on $det(W^+_\eufm{c}) $ and $R$ denotes the scalar curvature of the metric.

Throughout this paper, we restrict our attention to the case
that the gauge group is $SU(2)$ and the theory is coupled with
one hypermultiplet carrying the fundamental
representation (the $2$-dimensional representation).

\subsec{The Action Functional}

The topological action of the twisted $N=2$ super-Yang-Mills theory
coupled with the hypermultiplet is given by
\eqn\caaa{
S =-i \{Q, V_{T}\},
}
where
\eqn\caa{\eqalign{
V_{T} =
 \Fr{1}{h^2}\int d^4\!x\sqrt{g}\biggl[&
\chi^{\m\n}_a\left(H^a_{\m\n}
-i(F^{+a}_{\m\n} + q^\dagger \s_{\m\n}T^a q)\right)
-\Fr{1}{2} g^{\m\n}(D_\m\bar\phi)_a \l^a_{\n}
+\Fr{1}{8}[\phi,\bar\phi]_a \eta^a
\cr
&
 +\left(X_{\!\tilde q}^{\a}\psi_{\!q\a}
      + \psi_{\!\tilde{q}}^{\a}X_{\!q\a}\right)
 +{i}\left(q^\dagger_{\dot\a}
\bar\phi_{a}T^a\bar\psi_{\!\tilde{q}}^{\dot\a}
+\bar\psi_{\!q\dot\a} \bar\phi_{a}T^a q^{\dot\a}\right)
\biggl].\cr
}}
The supersymmetry transformation laws for the fields
in the adjoint representation are
\eqn\baa{\eqalign{
\hat\delta A_\m &= i\vr \lambda_\m,\cr
\hat\delta \lambda_\m &= -\vr D_\m \phi,\cr
\hat\delta \phi &= 0,\cr
}\qquad
\eqalign{
\hat\delta\chi_{\m\n} &= \vr H_{\m\n},\cr
\hat\delta \bar\phi &= i\vr\eta,\cr
}\qquad\eqalign{
\hat\delta H_{\m\n} &= i\vr[\phi,\chi_{\m\n}],\cr
\hat\delta\eta &= \vr[\phi,\bar\phi],\cr
}
}
where  $\hat\d(field) = -i\vr\{Q,field\}$.
The fields carrying the representation $R$ and $\tilde R$
transform as
\eqn\bag{
\eqalign{
\hat\delta q^{\dot\a}=
 &-\vr \bar{\psi}_{\!\tilde{q}}^{\dot{\alpha}},
  \cr
\hat\delta q_{\dot\a}^{\dagger}=
 &-\vr \bar{\psi}_{\!q \dot{\alpha}} ,
 \cr
}
\qquad\eqalign{
\hat\delta\bar{\psi}_{\!\tilde{q}}^{\dot{\alpha}}=
& -i\vr \phi^{a}T_a q^{\dot{\a}},
  \cr
\hat\delta \bar{\psi}_{\!q\dot{\alpha}}=
&i\vr q_{\dot\a}^{\dagger}\phi^{a}T_{a},
  \cr
}
}
and
\eqn\bbd{
\eqalign{
\hat\delta \psi_{\!q\a} &= -i\vr \s^{\m}{}_{\a\dot\a} D_\m q^{\dot\a}
                       +\vr X_{q\a},
\cr
\hat\delta X_{q\a}
&=i\vr\phi^a T_a\psi_{\!q\a}
-i\vr \s^{\m}{}_{\a\dot\a}D_\m \bar\psi_{\!\tilde{q}}^{\dot\a}
+\vr \s^{\m}{}_{\a\dot\a}\lambda_\m^a T_a q^{\dot\a},
\cr
\hat\delta \psi_{\!\tilde q}^{\a} &= i\vr D_\m q_{\dot\a}^\dagger
      \bar \s^{\m\dot\a\a} -\vr X_{\tilde{q}}^{\a},
\cr
\hat\delta X_{\tilde{q}}^{\a}
&=i\vr\psi_{\!\tilde{q}}^{\a}\phi^a T_a
-i\vr D_\m \bar\psi_{\!q\dot\a} \bar\s^{\m\dot\a\a}
+\vr q^\dagger_{\dot\a}\bar\s^{\m\dot\a\a} \lambda_\m^a T_a. \cr
}
}

The above transformation laws satisfy
\eqn\commutation{
(\hat\delta_\vr\hat\delta_{\vr^\pr} -\hat \delta_{\vr^\pr}\hat\delta_\vr)
(field) = T_\e(field),
}
where $T_\e(field)$ denotes the variation of a field under
a gauge transformation generated by an infinitesimal parameter
 $\e = -2i\vr\vr^\pr\cdot \phi$.
We introduce an additive quantum $U$-number called
the ghost number. The global supercharge $Q$ carries
$U =1$ and $V_T$ is designed to have $U=-1$ such
that the action has the zero-ghost number. So the theory
has the $U$-number symmetry at the classical level.
The $U$-number for the various fields are given by
\eqn\unumber{
\matrix{
&A_\m& \l_\m& \phi&\bar\phi&\eta&\chi_{\m\n}
& H_{\m\n}&q^{\dot{\a}}&q^\dagger_\a
& \bar\psi^{\dot\a}_{\!\tilde q}& \bar\psi_{\! q\dot{\a}}
&\psi_{\! q\a}&\psi^{\a}_{\!\tilde{q}}& X_{\! q\a}& X^\a_{\!\tilde{q}}\cr
&&&&&&&&&&&\cr&
0&1&2&-2&-1&-1&0&0&0&1&1&-1&-1&0&0\cr
}.
}

The topological action is given by\foot{We replaced  the complex
scalar fields $B$ and $\bar B$ of the physical theory
with $\Fr{i}{2\sqrt{2}}\phi$ and $i\sqrt{2}\bar\phi$.
This does not necessarily mean that $\phi$ and $-\bar\phi$
should be complex conjugate after twisting.}
\eqn\act{
\eqalign{
S = \Fr{1}{h^2}\int\!&d^4\!x\sqrt{g}\biggl[
\left( H^{\m\n}_a
  - \Fr{i}{2}(F^{+\m\n}_a
  +q^\dagger\bar\s^{\m\n}T_a q
)\right)
\left( H_{\m\n}^a
- \Fr{i}{2}(F^{+a}_{\m\n}
+ q^\dagger\bar\s_{\m\n}T^a q
)\right)
\cr
&
+\Fr{1}{4}\left(F^{+\m\n}_a
+q^\dagger\bar\s^{\m\n}T_a q\right)
\left(F^{+a}_{\m\n}
+q^\dagger\bar\s_{\m\n}T^a q\right)
-{2} X_{\!\tilde{q}}^{\a} X_{\!q\a}
\cr
&
+{i} X_{\!\tilde{q}}^{\a}\s^{\m}{}_{\a\dot\a}D_\m q^{\dot\a}
+{i}D_\m q^\dagger_{\dot\a}\bar\s^{\m\dot\a\a}X_{\!q\a}
-\Fr{1}{2}g^{\m\n}(D_\m\bar\phi)_a(D_\n\phi)^a
+\Fr{1}{8}[\phi,\bar\phi]_a[\phi,\bar\phi]^a
\phantom{\biggr]}
\cr
&
-i\chi^{\m\n}_a[\phi,\chi_{\m\n}]^a
+\chi^{\m\n}_a(d_A\l)^{+a}_{\m\n}
+\Fr{i}{2}g^{\m\n}(D_\m\eta)_a\l^a_\n
-\Fr{i}{2}g^{\m\n}[\l_\m,\bar\phi]_a \l_\n^a
+\Fr{i}{8}[\phi,\eta]_a \eta^a
\cr
&
-{i}\chi^{\m\n}_{a}\bar\psi_{\!q}\bar\s_{\m\n}T^a q
+{i}\chi^{\m\n}_{a}
  q^\dagger\bar\s_{\m\n}T^a \bar\psi_{\!\tilde{q}}
-{i}D_\m\bar\psi_{\!q\dot\a}\bar\s^{\m\dot\a\a}\psi_{\!q\a}
-{i}\psi_{\!\tilde{q}}^\a\s^{\m}{}_{\a\dot\a}
 D_\m\bar\psi_{\!\tilde{q}}^{\dot\a}
\phantom{\biggr]}\cr
&
+2 i\psi_{\!\tilde{q}}^\a\phi_a T^a \psi_{\!q\a}
-2i\bar\psi_{\!q\dot\a}\bar\phi_a T^a \bar\psi_{\!\tilde{q}}^{\dot\a}
-q^\dagger_{\dot\a}\l_{\m{a}} T^a \bar\s^{\m\dot\a\a}\psi_{\!q\a}
-\psi_{\!\tilde{q}}^\a \s^{\m}{}_{\a\dot\a}\l_{\m{a}} T^a
q^{\dot\a}. \phantom{\biggr]}\cr
&
+ q^\dagger_{\dot\a}\eta_a T^a \bar\psi_{\!\tilde q}^{\dot\a}
- \bar\psi_{\!q\dot\a}\eta_a T^a q^{\dot\a}
- q^\dagger_{\dot\a}T^a T^b
\left(\phi_a\bar\phi_b + \phi_b\bar\phi_a\right)q^{\dot\a}
\biggr].\cr
}
}
After integrating out  the auxiliary fields
$H_{\m\n}$, $X_{\!\tilde{q}}^{\a}$ and $X_{\!q\a}$,
we have
\eqn\action{\eqalign{
S=
\Fr{1}{h^2}&\int\!d^4\!x\sqrt{g}\biggl[
\Fr{1}{4}F^{+\m\n}_a F^{+a}_{\m\n}
-\Fr{1}{2}p^{+}_{\m\n}q^\dagger\bar\s^{\m\n}q
+\Fr{1}{2}g^{\m\n}D_\m q^\dagger_{\dot\a} D_\n q^{\dot\a}
\cr
&
+ \Fr{1}{4}(q^\dagger\bar\s^{\m\n} T_a q )
                  (q^\dagger\bar\s_{\m\n} T^a q)
+\Fr{1}{8}R (q^\dagger_{\dot\a}q^{\dot\a})
-\Fr{1}{2}g^{\m\n}(D_\m\bar\phi)_a(D_\n\phi)^a
+\Fr{1}{8}[\phi,\bar\phi]_a[\phi,\bar\phi]^a
\phantom{\biggr]}
\cr
&
-i\chi^{\m\n}_a[\phi,\chi_{\m\n}]^a
+\chi^{\m\n}_a(d_A\l)^{+a}_{\m\n}
+\Fr{i}{2}g^{\m\n}(D_\m\eta)_a\l^a_\n
-\Fr{i}{2}g^{\m\n}[\l_\m,\bar\phi]_a \l_\n^a
+\Fr{i}{8}[\phi,\eta]_a \eta^a
\cr
&
-{i}\chi^{\m\n}_{a}\bar\psi_{\!q}\bar\s_{\m\n}T^a q
+{i}\chi^{\m\n}_{a}
  q^\dagger\bar\s_{\m\n}T^a \bar\psi_{\!\tilde{q}}
-{i}D_\m\bar\psi_{\!q\dot\a}\bar\s^{\m\dot\a\a}\psi_{\!q\a}
-{i}\psi_{\!\tilde{q}}^\a\s^{\m}{}_{\a\dot\a}
 D_\m\bar\psi_{\!\tilde{q}}^{\dot\a}
\phantom{\biggr]}\cr
&
+2 i\psi_{\!\tilde{q}}^\a\phi_a T^a \psi_{\!q\a}
-{2}i \bar\psi_{\!q\dot\a}\bar\phi_a T^a \bar\psi_{\!\tilde{q}}^{\dot\a}
+q^\dagger_{\dot\a}\l_{\m{a}} T^a \bar\s^{\m\dot\a\a}\psi_{\!q\a}
+\psi_{\!\tilde{q}}^\a \s^{\m}{}_{\a\dot\a}\l_{\m{a}} T^a
q^{\dot\a}. \phantom{\biggr]}\cr
&
+ q^\dagger_{\dot\a}\eta_a T^a \bar\psi_{\!\tilde q}^{\dot\a}
- \bar\psi_{\!q\dot\a}\eta_a T^a q^{\dot\a}
- q^\dagger_{\dot\a}T^a T^b \left(\phi_a\bar\phi_b +
 \phi_b\bar\phi_a\right)q^{\dot\a}
\biggr].\cr
}
}
The above action is $Q$ invariant after changing the transformation
law as
\eqn\caj{\eqalign{
\hat\delta \chi^{a}_{\m\n}
&= \Fr{i}{2}\vr(  F^{+a}_{\m\n}
        +q^\dagger\bar\s_{\m\n}T^a q),
\cr
\hat\delta \psi_{\!q\a}
&=- \Fr{i}{2}\vr \s^{\m}{}_{\a\dot\a}
  D_\m q^{\dot\a},
\cr
\hat\delta \psi_{\!\tilde q}^{\a}
&=\Fr{i}{2}\vr D_\m q_{\dot\a}^\dagger \bar\s^{\m\dot\a\a}.
\cr
}
}

By the fixed point theorem of Witten the path integral is localized to the
locus of the fixed point of the global supersymmetry,
modulo gauge symmetry.
The important fixed points  are
$\delta \chi_{\m\n} = \delta \psi_{\!q\a}= 0$;
\eqn\monopole{
F^{+a}_{\m\n} +q^\dagger\bar\s_{\m\n}T^a q =0,
\qquad
\s^{\m} D_\m q =0,
}
which is the non-abelian version of the Seiberg-Witten monopoles.
Note that the curvature $F_{\m\n}$ is the curvature of the bundle $E$
(or $P$), while $q \in \G(W^+_\eufm{c}\otimes E)$ is the section
of $W^+_\eufm{c}\otimes E$. The Dirac operator
$\s^\m D_\m :\G(W^+_\eufm{c}\otimes E)\rightarrow
\G(W^-_\eufm{c}\otimes E)$
is the Dirac operator for the $spin^c$ bundle twisted by $E$.\foot{
In particular, if we consider the
$U(1)$ gauge theory, $G=U(1)$ such that $E$ is a line bundle, the
equation \monopole\ is not identical to the Seiberg-Witten monopole
equation.
The only special property for the $U(1)$ theory is that if the first Chern
class $c_1(E)$ is integral the tensor product bundle
$W^+_\eufm{c}\otimes E$ becomes another $spin^c$ bundle
$W^+_\eufm{c^\pr}$ defined by the new $spin^c$ structure
$\eufm{c^\pr} = \eufm{c} + 2 c_1(E)$.
Thus the equation that can be obtained by twisting the $N=2$
super-Maxwell theory coupled
with hypermultiplet is a perturbed Seiberg-Witten equation rather
than the original equation.
}
Note also that the curvature $F$ is a $ad(P)$-valued $2$-form
(or trace-free endomorphism $End(E) = E\otimes \tilde E$
valued $2$-form). Since
$q \in \G(W^+_\eufm{c}\otimes E)$ and
${q}^\dagger \in \G(\overline{W}^+_\eufm{c}\otimes \tilde{E})$,
the product $q\otimes  q^\dagger$ lies in
$$
W^+\otimes_\eufm{c} E\otimes \overline{W}^+_\eufm{c}\otimes \tilde{E}
\sim \O^0(End(E))\oplus\O^2_+(End(E)),
$$
where $\O^0(End(E))$ and
$\O^2_+(End(E))$ denote the spaces of $End(E)$-valued zero-forms
and $End(E)$-valued self-dual-two-forms respectively.
An equivalent description can be obtained by examining the
semi-classical limit $h^2 \rightarrow 0$, which is exact.
The relevant bosonic part of the action can be written
as
\eqn\bosonic{\eqalign{
\Fr{1}{h^2}\int d^4\!x\sqrt{g}&\left(\Fr{1}{4}|s|^2 + \Fr{1}{2} |k|^2\right)
\cr
&=\Fr{1}{h^2}\int\!d^4\!x\sqrt{g}\biggr(
\Fr{1}{4}F^{+\m\n}_a F^{+a}_{\m\n}
-\Fr{1}{2}p^{+}_{\m\n} q^\dagger\bar\s^{\m\n} q
+\Fr{1}{2}g^{\m\n}D_\m q^\dagger_{\dot\a} D_\n q^{\dot\a}
\cr
&\qquad\qquad\qquad
+ \Fr{1}{4}(q^\dagger\bar\s^{\m\n} T_a q )
                  (q^\dagger\bar\s_{\m\n} T^a q)
+\Fr{1}{8}R (q^\dagger_{\dot\a}q^{\dot\a})
\biggl),
}}
 where $s = F^{+a}_{\m\n} +q^\dagger\bar\s_{\m\n}T^a q$ and
$k= \s^{\m} D_\m q$. Thus, the path integral has the dominant
contributions from the solutions of \monopole.

We also have another fixed point equation
$\delta \l_\m =\delta \eta =0$;
\eqn\redu{
D_\m\phi =0,\qquad
[\phi,\bar\phi] =0.
}
A connection is reducible if there exists non-zero solution
$\phi$ of $D_\m \phi = 0$.

The virtual (or formal) dimension of the moduli space $\CM(\eufm{c},k)$,
the space
of solutions of \monopole\ in $\CA\times \G(W^+_\eufm{c}\otimes E)$
modulo the gauge symmetry, is
\eqn\mma{\eqalign{
{dim}\; \CM(k,\eufm{c}) &=
index\left( d_A^+\oplus d_A^*\right)
+ 2{index}\left(\Fs{D}^E_{\eufm{c}}\right)=
{dim}\; \CM(k) + 2{index}\left(\Fs{D}^E_{\eufm{c}}\right)
\cr
&= 8k -\Fr{3}{2}(\chi + \s)
-2k+ \Fr{1}{2}(\eufm{c}\cdot\eufm{c} -\s),
}}
where $\CM(k)$ denotes the moduli space of $SU(2)$ ASD
connection with the instanton number  $k$
and $'\cdot'$ denotes the intersection pairing.
We will use the following notations;
\eqn\coon{
2d(\eufm{c},k)=\hbox{dim}\; \CM(k,\eufm{c}),\quad
2d(k) = \hbox{dim}\; \CM(k),\quad
d_0(\eufm{c}, k) = d(\eufm{c},k)-d(k) = index\; \Fs{D}^E_{\eufm{c}}.
}
With the choice of $b_2^+ = 1 + 2a$ for a positive integer $a$,
both dimensions are even.
We would like to remind the reader that the net $U$-number violation
$\triangle U$ in the path integral measure due to the
fermionic zero-modes equals  the virtual dimension of the moduli space
$\CM(k,\eufm{c})$ \HPPb.
The virtual dimension of $\CM(\eufm{c},k)$ becomes the real dimension
if there are no fermionic zero-modes except those of
$\l$ and $( \bar\psi_{\!\tilde q\dot{\a}}, \bar\psi^{\dot\a}_{\tilde q})$
which span the tangent space over $\CM(\eufm{c},k)$.
However, such an ideal situation will hardly be the case.

The actions \act\ and \action\ are also invariant under the global
scaling of the metric if the scaling dimensions of the
various fields are assigned as
\eqn\snumber{
\matrix{
&A_\m& \l_\m& \phi&\bar\phi&\eta&\chi_{\m\n}
& H_{\m\n}&q^{\dot{\a}}&q^\dagger_\a
& \bar\psi^{\dot\a}_{\! \tilde q}& \bar\psi_{\! q\dot{\a}}
&\psi_{\! q\a}&\psi^{\a}_{\!\tilde{q}}& X_{\! q\a}& X^\a_{\!\tilde{q}}\cr
&&&&&&&&&&&\cr&
1&1&0&2&2&2&2&1&1&1&1&2&2&2&2\cr
}.
}
If  an operator $O$ has the scaling dimension $n$,
the integral $\int d^4\! x\sqrt{g} O$ scales as
$t^{4-n}$ under $g\rightarrow t g$.
The transformation laws \baa\bag\ and \bbd\ also preserve
the scaling dimensions.

The crucial property of any cohomological field theory
is that the energy-momentum tensor should be a
$Q$-commutator,
\eqn\ema{
\d S = \Fr{1}{2}\int_X \sqrt{g}\d g^{\m\n} T_{\m\n},
\qquad T_{\m\n} = \{Q,\l_{\m\n}\}.
}
This immediately follows from the relation \caaa\
if the variation operator $\d/\d g^{\m\n}$ commutes
with $Q$ in off shell. The only subtlety comes from  the
fields which are subject to the self-duality condition
that the variation of the metric should be accompanied
by the variations of the fields to preserve the self-duality.
Since the algebra \baa\ is closed in off shell it can be
guaranteed.
Most of the important properties of the
topological theory can be derived from the property \ema.
In particular, the topological invariance of the
suitable correlation function is based on the property.
The only obstruction is the possible metric dependency
of the path integral measure. For a more detailed
discussion on those properties, we refer the reader to
\WittenA.

\subsec{The $\CG$-Equivariant Cohomology}

In this subsection, we briefly describe the relation between
the twisted supersymmetry and the equivariant cohomology
\WittenE\REVb.

Consider the space
$\CA\times \G(W^+_\eufm{c}\otimes  E )$.
The group $\CG =Map(X,G)$ of the
gauge transformation acts on $\CA$ in the usual way and on
$\G(W^+_\eufm{c}\otimes E)$
according to the representation $R$ and $\tilde R$ of $G$.
Let $Lie(\CG)$ be the Lie algebra of $\CG$.
The $\CG$ action
on $\CA\times \G(W^+_\eufm{c}\otimes E )$
is generated by vector fields  $V_{\rm a}$,
where we pick an orthonormal basis $\CT_{\rm a}$ of $Lie(\CG)$.
Let $\hbox{Fun($Lie(\CG)$)}$ be the algebra  of polynomial functions,
generated by $\phi^{\rm a}$ with degree $2$, on $Lie(\CG)$.

Now, one can formally define the (infinite dimensional)
$\CG$-equivariant de Rham cohomology.
Let $\O^*(\CA\times \G(W^+_\eufm{c}\otimes E ))$
be the de Rham complex on
$\CA\times \G(W^+_\eufm{c}\otimes E )$.
The equivariant de Rham complex is defined by
\eqn\maa{
\O^*_\CG(\CA\times \G(W^+_\eufm{c}\otimes  E)
=\left(\O^*(\CA\times \G(W^+_\eufm{c}\otimes E ))
\otimes Fun(Lie(\CG))\right)^\CG.
}
The associated differential operator  $\d$
can be formally  represented
as
\eqn\mab{
\d = -\int  d^4\!x \sqrt{g}\P^I(x)\Fr{\d}{\d A^{I}(x)}
+i\int d^4\!x \sqrt{g}V(\phi(x))  \Fr{\d}{\d \P^{I}(x)},
}
where $A^I$ denote collectively the local coordinates on
$\CA\times \G(W^+_\eufm{c}\otimes E)$
and $\P^I$ denote the basis of
the cotangent space.
We have
\eqn\mac{
\d^2 = -i\int d^4x\sqrt{g} \phi^{\rm a}(x)\CL_{\rm a}(x),
}
where  $\CL_{\rm a}$ is the Lie derivative
with respect to $V_{\rm a}$. Thus, $\d^2 = 0$
on the $\CG$-invariant subspace
$\O^*_\CG(\CA\times \G(W^+_\eufm{c}\otimes E ))$
of  $\O^*(\CA\times \G(W^+_\eufm{c}
\otimes E))\otimes \hbox{Fun}(Lie(\CG))$.
The $\CG$-equivariant de Rham cohomology
$H^*_\CG(\CA\times \G(W^+_\eufm{c}
\otimes E))$ is defined as
the pairs $(\O^*_\CG(\CA\times
\G(W^+_\eufm{c}\otimes E )),\;\d)$.

The basic supersymmetry algebra
\eqn\mad{
\eqalign{
\hat\delta A_\m =& + i\vr \lambda_\m,\cr
\hat\delta q^{\dot\a}=
 &-\vr \bar{\psi}_{\!\tilde{q}}^{\dot{\alpha}},
  \cr
\hat\delta q_{\a}^{\dagger}=
 &-\vr \bar{\psi}_{\!q \dot{\alpha}} ,
 \cr
}
\qquad\eqalign{
\hat\delta \lambda_\m =& -\vr D_\m \phi,\cr
\hat\delta\bar{\psi}_{\!\tilde{q}}^{\dot{\alpha}}=
& -i\vr \phi^{a}T_a q^{\dot{\a}},
  \cr
\hat\delta \bar{\psi}_{\!q\dot{\alpha}}=
&+i\vr q_{\dot\a}^{\dagger}\phi^{a}T_{a},
  \cr
}\qquad
\hat\delta \phi = 0,
}
suggests that
the twisted supercharge of the theory without the mass term
can be interpreted
as the generator of the $\CG$-equivariant de Rham cohomology
$H^*_\CG(\CA\times \G(W^+_\eufm{c}\otimes E))$.
The relation
\eqn\mae{
(\hat\delta_\vr\hat\delta_{\vr^\pr} -\hat \delta_{\vr^\pr}\hat\delta_\vr)
(field) = T_\e(field),\quad\hbox{where}\quad
\e = -2i\vr\vr^\pr\cdot \phi,
}
corresponds to  the property \mac.

Similarly to the TYM theory,
one can interpret the twisted supercharge
$Q$ of TQCD as the $\CG$-equivariant cohomology operator.
Then, the topological
action can also be interpreted as a
certain Mathai-Quillen representative
of a universal Thom class \AJ\ based on the non-abelian version
of Seiberg-Witten equations.  For such a construction of the similar
topological theory, we refer the reader to the paper  \LM.

\subsec{The Observable and the Correlation Function}

In the simply connected Riemann manifold,
the second cohomology
class determines the essential
cohomological data. Picking a $2$-dimensional homology
class $\S \in H_2(X;\BZ)$ which is Poincar\`{e} dual to
$v\in H^2(X;\BZ)$, one defines the associated
topological observable
\eqn\mba{
\hat v = \Fr{1}{4\pi^2}\int_\S \tr(i \phi F + \Fr{1}{2}\l\wedge\l),
}
which carries the $U$-number $2$.
This observable $\hat v$ defines
a $2$-dimensional $\CG$-equivariant cohomology class, i.e.,
$ \hat v \in H^2_\CG(\CA\times \G(W^+_\eufm{c}\otimes E))$.
The $Q$-cohomology class of $\hat v$ depends only on the
homology class of $\S$. One also has the topological
observable
\eqn\mbab{
\hat u = -\Fr{1}{8\pi^2}\tr \phi^2,
}
carrying the $U$-number $4$ and depending only
on $H_0(X;\BZ)$. The observable $\hat u$ defines
a $4$-dimensional class $H^4_\CG$.
TQCD has no additional non-trivial
topological observables beyond those of the TYM theory.
This may be an indication that the theory would have
no new  differential-topological information.

Now we consider the topological correlation function
\eqn\mbc{
\left< {\hat v}^r \hat u^s \right>_{TQCD}
= \Fr{1}{\hbox{vol}(\CG)}\int \CD Y e^{-S}\cdot
{\hat v}^r\hat u^s.
}
Due to the ghost number anomaly in the path integral
measure  the above topological amplitude vanishes
unless  $2r + 4s$ is identical to the {\it formal dimension}
$2d(\eufm{c},k)$ of the moduli space $\CM(\eufm{c},k)$.
If we consider, presumably, the favorable condition that
the formal dimension of the moduli space is the actual
dimension,  the path integral
reduces to an integration of the wedge product of
differential forms on the moduli space $\CM(\eufm{c},k)$.
This can be seen by  both the $Q$-fixed point theorem
and the semi-classical analysis.
The differential form is given by the restriction and the reduction
$\hat v_0$ of $\hat v$ to the moduli space $\CM(\eufm{c},k)$.
That is, $\hat v_0$ defines an element of the de Rham cohomology
class on $\CM(\eufm{c},k)$.\foot{Of course, the above
cohomological definition can be mathematically meaningful
after some suitable compactification of the moduli space is
understood. Here the above argument is
just a formal cohomological interpretation of
the path integral.
{\it Whatever properties the moduli space has,
the path integral, as we shall see,
gives a perfectly concrete formula for those invariants}.
See \WittenA.}
The standard recipe similar to the TYM theory leads that
$\hat v_0$ can be obtained by replacing $\l$ by its zero-modes,
$F$ by the $Q$-fixed point value and $\phi$ by $<\phi>$.

The integration over $\bar\phi$ in \action\
gives a delta-function constraint
\eqn\rema{
\Fr{1}{2}g^{\m\n}D_\m D_\n \phi^a+ \Fr{i}{2} g^{\m\n}[\l_\m,\l_\n]^a
+q^\dagger_{\dot{\a}}(T^a T^b +T^b T^a)q^{\dot\a} \phi^b
-2i\bar\psi_{\!q\dot{\a}}T^a\bar\psi^{\dot\a}_{\!\tilde q} = 0.
}
The $\eta$ equation of motion of \action\ gives
\eqn\etaaa{
\Fr{1}{2}g^{\m\n}D_\m \l_\n
+ i q^\dagger_{\dot\a}T^a \bar\p^{\dot\a}_{\!\tilde q}
+ i \bar\p_{\! q\dot{a}}T^a q^{\dot\a} =0,
}
which expresses the fact that the zero-modes of
$(\l, \bar\p^{\dot\a}_{\!\tilde q}, i \bar\p_{\! q\dot{a}})$
are orthogonal to the gauge variation.
The $\bar\phi$ equation of motion \rema\ is just
the supersymmetry transformation of \etaaa.
We can write \rema\ as
\eqn\remab{
\left(D^\m D_\m\delta^{a}_{b} + 2q^\dagger_{\dot{\a}}
(T^a T^b + T^b T^a)q^{\dot\a}\right)\phi^b
= -i [\l^\m,\l_\m]^a + 4i\bar\psi_{\!q\dot{\a}}T^a
\bar\psi^{\dot\a}_{\!\tilde q}.
}
On the other hand, the theory without coupling to the
matter leads to
\eqn\remabc{
\left(D^\m D_\m \right)\phi  = -i [\l^\m,\l_\m] .
}

Furthermore, we can substitute the vacuum expectation
value $<\phi>$ of $\phi$ by
\eqn\remabb{
<\phi^a> = - i\int_X d^4y\sqrt{g}\, G^{a b}(x,y)
\left( [\l^\m(y),\l_\m(y)]^b - 4\bar\psi_{\!q\dot{\a}}(y)T^b
\bar\psi^{\dot\a}_{\!\tilde q}(y)\right),
}
where
\eqn\rra{
\left(D^\m D_\m\delta_{a b}+2
q^\dagger_{\dot{\a}}(T_a T_b + T_b T_a)q^{\dot\a}
\right)
G^{a b}(x,y)
= \d^{ab}\delta^4(x-y),
}
provided that we replace  $q^{\dot\a}$,
$q^\dagger_{\dot{\a}}$ by the non-abelian Seiberg-Witten
monopole \monopole\ and $\l_\m$, $\bar\psi_{\!q\dot{\a}}$
and $\bar\psi^{\dot\a}_{\!\tilde q}$ by their zero-modes
which represent the tangent vectors of the moduli space
$\CM(k,\eufm{c})$.
As the standard recipe of the TYM theory, we can replace
$\phi$ with \remabb\ whenever it appears in the topological
observables.\foot{In TYM theory, the replacement of $\phi$
with $\left<\phi\right>$ is the procedure for recovering the
universal bundle construction of the cohomology
class on $\CA/\CG$ \DK\REVa\REVb. The formula \remabb\
implies that such a replacement in \mba\ and \mbab\
will recover the analogous universal bundle construction
for the extended space.}
This explains one of the mysteries of  TQCD
that no new and non-trivial observables are introduced due
to the hypermultiplet.

\newsec{The Massive TQCD}

In the previous
sections we only considered the theory with massless hyper-multiplet
and its topological twisting.   If we twist the $N=2$ supersymmetric
Yang-Mills theory coupled with massive hypermultiplet,
a remarkable
thing happens. We will show that the correlation
function can be expressed by the sum of the contributions due to
the Donaldson invariants and the Seiberg-Witten  invariants.

\subsec{The Massive Hypermultiplet}

In the $N=2$ supersymmetric QCD, the hypermultiplets
can have the bare mass term which is
invariant under the $N=2$ supersymmetry.
We will always consider  the theory with one hypermultiplet.
In the on-shell action, the mass term can be written as\foot{
In terms of $N=1$ superspace notation,
the action functional for hypermultiplet is given by
$
W=\sqrt{2} \tilde Q_h \Phi Q_h + m \tilde Q_h Q_h,
$
where $Q_h$ and $\tilde Q_h$ are chiral superfields carrying
a representation $R$ and its conjugate $\tilde R$, respectively,
$\Phi$ is the $N=1$ chiral multiplet carrying the adjoint
representation and $m$ is the bare mass
for hypermultiplet.
The untwisted action functional we are using is just the expansion
of $W$ in terms of the component fields. We follow the conventions
in our previous paper \HPPa.}
\eqn\mca{
S_{mass} = \int d^4 x \biggr[-m^2 q^\dagger_i q^i
 - \sqrt{2} m q^\dagger_i B^a T_a q^i
 + \sqrt{2} m q^\dagger_i \bar B^a T_a q^i
 -m\bar\p^{\dot{\a}}_{\tilde{q}}\bar\p_{q\dot\a}
 -m\p^\a_{\tilde{q}}\p_{q\a}\biggl].
}
Adding the mass term leads to the following on-shell
supersymmetry transformation of the hypermultiplet
\eqn\mcb{
\eqalign{
\delta q^{i}=
&-\sqrt{2}\xi^{\alpha i}\psi_{q\alpha}
+\sqrt{2}\bar{\xi}_{\dot{\alpha}}{}^{i}
\bar{\psi}_{\tilde{q}}^{\dot{\alpha}},
   \cr
\delta\bar{\psi}_{\tilde{q}}^{\dot{\alpha}}=
&-\sqrt{2} i\bar{\sigma}^{m\dot{\alpha}\alpha}D_{m}q^{i}\xi_{\alpha i}
 +2T_{a}q^{i}B^{a}\bar{\xi}^{\dot{\alpha}}{}_{i}
 +\sqrt{2}m q^{i}\bar\xi^{\dot\a}{}_{i} ,
 \cr
  \delta\psi_{q\alpha}=
&-\sqrt{2}i\sigma^{m}_{\alpha\dot{\alpha}}
   D_{m}q^{i}\bar{\xi}^{\dot{\alpha}}{}_{i}
   -2T_{a}q^{i}\bar{B}^{a}\xi_{\alpha i}
   +\sqrt{2}m q^{i}\xi_{\a i} ,
  \cr
}
}
and of the conjugate fields
\eqn\mcc{\eqalign{
\delta q_{i}^{\dagger}=
&-\sqrt{2}\bar{\psi}_{q \dot{\alpha}}
\bar{\xi}^{\dot{\alpha}}{}_{i}
   -\sqrt{2}\psi_{\tilde{q}}^{\alpha}\xi_{\alpha i},
   \cr
\delta \bar{\psi}_{q\dot{\alpha}}=
&\sqrt{2}i\xi^{\alpha i}D_{m}q_{i}^{\dagger}
\sigma^{m}_{\alpha\dot{\alpha}}
 -2 \bar{\xi}_{\dot{\alpha}}{}^{i}q_{i}^{\dagger}B^{a}T_{a}
 -\sqrt{2}m \bar\xi_{\dot\a}{}^{i} q^\dagger_{i} ,
 \cr
\delta\psi_{\tilde{q}}^{\alpha}=
&-\sqrt{2}i\bar{\xi}_{\dot{\alpha}}^{i}
     D_{m}q_{i}^{\dagger}\bar{\sigma}^{m\dot{\alpha}\alpha}
  -2\xi^{\alpha i}\bar{B}^{a}q_{i}^{\dagger}T_{a}
  +\sqrt{2}m \xi^{\a i}q^\dagger_{i}. \cr
}}

If we twist the above supersymmetry, there appear several problems
concerning the mass term of the hypermultiplet. We first twist
the supersymmetry transformation laws of the
hypermultiplet following the recipe of our previous paper \HPPa.
The twisted transformation laws of the
hypermultiplet
are given by
\eqn\mcd{
\eqalign{
\hat\delta_m q^{\dot\a}=
 &-\vr \bar{\psi}_{\!\tilde{q}}^{\dot{\alpha}},
  \cr
\hat\delta_m q_{\a}^{\dagger}=
 &-\vr \bar{\psi}_{\!q \dot{\alpha}} ,
 \cr
}
\qquad\eqalign{
\hat\delta_m\bar{\psi}_{\!\tilde{q}}^{\dot{\alpha}}=
& -i\vr \phi^{a}T_a q^{\dot{\a}}
  -\vr m q^{\dot\a},
  \cr
\hat\delta_m \bar{\psi}_{\!q\dot{\alpha}}=
&i\vr q_{\dot\a}^{\dagger}\phi^{a}T_{a}
 +\vr m q^\dagger_{\dot\a},
  \cr
}
}
and
\eqn\mcda{
\eqalign{
\hat\delta_m \psi_{\!\tilde q}^{\a}
    &= i\vr D_\m q_{\dot\a}^\dagger
          \bar \s^{\m\dot\a\a} -\vr X_{\tilde{q}}^{\a},
 \cr
\hat\delta_m X_{\tilde{q}}^{\a}
    &=i\vr\psi_{\!\tilde{q}}^{\a}\phi^a T_a
        -i\vr D_\m \bar\psi_{\!q\dot\a} \bar\s^{\m\dot\a\a}
        +\vr q^\dagger_{\dot\a}\bar\s^{\m\dot\a\a} \lambda_\m^a T_a
        +\vr m \psi_{\!\tilde{q}}^{\a},
\cr
\hat\delta_m \psi_{\!q\a}
    &= -i\vr \s^{\m}{}_{\a\dot\a} D_\m q^{\dot\a}
           +\vr X_{q\a},
\cr
\hat\delta_m X_{q\a}
     &=i\vr\phi^a T_a\psi_{\!q\a}
         -i\vr \s^{\m}{}_{\a\dot\a}D_\m \bar\psi_{\!\tilde{q}}^{\dot\a}
          +\vr \s^{\m}{}_{\a\dot\a}\lambda_\m^a T_a q^{\dot\a}
         +\vr m \psi_{\!q\a},
}
}
while the transformation laws for the $N=2$ vector multiplet
remains unchanged. Note that the twisted algebra \mcd\
and \mcda\  closed in off shell.
Due to the new terms proportional to the mass $m$, the commutator
of the supersymmetry is no longer the gauge transformation generated
by $\phi$. Furthermore, the above twisted supersymmetry
does not preserve the $U$-number. Note, however, the global scaling
dimensions \snumber\ are preserved if the scaling dimension
of $m$ is $0$.

Due to the new transformation laws, the action functional has
the following additional terms;
\eqn\acadd{
S^\pr =-i\{Q_m,V_T\}= S + \Fr{1}{h^2}\int_X d^4 x \sqrt{g}
\left(2 im q^\dagger_{\dot\a}  \bar\phi^aT_a q^{\dot\a}
+m\psi^{\a}_{\!\tilde{q}}\psi_{\!q\a}\right).
}
Note that the additional terms carry the $U$-number $-2$
while $S^\pr$ maintains the global scale invariance.
The property that the energy-momentum tensor
is a $Q_m$ commutator remains unchanged.
\eqn\emb{
T^\pr_{\m\n} = \{Q_m,\l_{\m\n}\}.
}

Now we have two problems;

(1) The modified transformation laws \mcd\ and \mcda\ for
the hypermultiplet
break
the basic commutation relation \mae\ due to the mass term;
\eqn\commut{
(\hat\delta_{m\vr}\hat\delta_{m\vr^\pr} -\hat \delta_{\vr^\pr}\hat\delta_\vr)
(hypermultiplet) = T_\e(hypermultiplet) -2\vr\vr^\pr\cdot m ,
}
where $\e = -2i\vr\vr^\pr\cdot \phi$ as before.
Furthermore, they do not preserve the $U$-number.
Note, however, that if we assign the $U$-number $2$ to $m$
the $U$-number is preserved.

(2) The modified action \acadd\ does not contain the full mass terms
of the hypermultiplet.

We will temporarily ignore the problem (1) which will be
resolved in Sect.~3.5.

Another very important effect is that the $\bar\phi$ equation of motion is
changed from \rema\ to
\eqn\mcdd{\eqalign{
\Fr{1}{2}g^{\m\n}D_\m D_\n \phi^a+ \Fr{i}{2} g^{\m\n}[\l_\m,\l_\n]^a
+q^\dagger_{\dot{\a}}(T^a T^b +T^b T^a)q^{\dot\a} \phi^b
-2i\bar\psi_{\!q\dot{\a}}T^a\bar\psi^{\dot\a}_{\!\tilde q}
-2im q^\dagger_{\dot\a}T^a q^{\dot\a} = 0.
\cr
\left(D^\m D_\m\delta^{a}_{b} + 2q^\dagger_{\dot{\a}}
(T^a T^b + T^b T^a)q^{\dot\a}\right)\phi^b
= -i [\l^\m,\l_\m]^a + 4i\bar\psi_{\!q\dot{\a}}T^a
\bar\psi^{\dot\a}_{\!\tilde q} + 4i m q^\dagger_{\dot\a}T^a q^{\dot\a},
\cr
}
}
and the equation \remabb  is changed accordingly.

We have the same form of topological observable
$\hat v$ in \mba\
which carries the $U$-number $2$ and is invariant under the
global scaling of the metric.
The $Q$ or $Q_m$-cohomology class of $\hat v$ depends only on the
homology class of $\S$. The same is true for $\hat u$ in \mbab.
However, they are effectively different from the massless case
due to the difference between \remab\ and \mcdd. We shall see
that the key simplification occurs due to the extra term in \mcdd.
The topological correlation function of the massive theory
\eqn\mbcr{
\left< {\hat v}^r \hat u^s \right>_{TQCD,m}
= \Fr{1}{\hbox{vol}(\CG)}\int \CD Y e^{-S^\pr}\cdot
{\hat v}^r\hat u^s,
}
also has the same $U$-number anomaly cancellation
laws, $r + 2s = d(\eufm{c},k)$.

\subsec{The $Q_m$-Fixed Points}

The effect of introducing the bare mass to the hypermultiplet
can be most easily seen by checking the fixed point equations
for the new global supercharge $Q_m$.
In addition to  the fixed point equations \monopole\
and \redu\ of the original
supersymmetry, we have new fixed point equations
\eqn\new{
\left\{
\eqalign{
\hat\d_m \bar\p^{\dot\a}_{\!\tilde{q}} =0,\cr
\hat\d_m \bar\p_{\!q\dot{\a}}=0,\cr
}
\right.
\Longrightarrow
\left\{
\eqalign{
m q^{\dot\a}+ i \phi_a T^a q^{\dot{\a}} = 0,\cr
m q^\dagger_{\dot\a}+i q_{\dot\a}^{\dagger}\phi^{a}T_{a}=0,\cr
}\right.
}
If we collect the other
$Q_m$-fixed point equations,
$\hat\delta_m \chi_{\m\n} = \hat\delta_m \psi_{\!q\a}=
\hat \delta_m \l_\m =\hat\delta_m \eta =0$,
we have
\eqn\mfixed{\left\{\eqalign{
F^{+a}_{\m\n} +q^\dagger\bar\s_{\m\n}T^a q =0,\cr
\s^{\m} D_\m q =0,\cr
}\right.\qquad
\left\{
\eqalign{
D_\m\phi =0,\cr
[\phi,\bar\phi] =0,\cr
}\right.
}
The first pair of the equations say that the fixed point
locus is the moduli space $\CM(\eufm{c},k)$.
The second pair of the equations means that
$\phi$ is zero at the fixed point locus if the
connection is irreducible (the  gauge symmetry
is unbroken) and $\phi$ is non-zero if the connection
is reducible (the gauge symmetry is broken down to
$U(1)$).

This is the judicious moment to
study the new fixed point
equation \new.
One obvious fixed point is $q^\dagger_{\dot\a}=q^{\dot\a}=0$,
which will be called  branch (i).
In this branch, the fixed point equations reduce
to those of the TYM theory;
\eqn\tymfix{
F^{+a}_{\m\n} =0,\qquad
D_\m\phi =0,\qquad
[\phi,\bar\phi] =0.
}
Thus the fixed point locus is the moduli
space of irreducible ASD connections
for $\phi =0$.\foot{For the manifold with $b_2^+ > 1$,
there are no reducible ASD connections
for a generic choice of the metric as well as
for a smooth path joining two generic metrics.
We will assume that the moduli space $\CM(k)$
is connected.
}
Note also that the fixed point equation \new\
reduces to $q^\dagger_{\dot\a}=q^{\dot\a}=0$
if $\phi = 0$. {\it So, whenever gauge symmetry
is unbroken, the fixed point locus of $Q_m$
is the moduli space $\CM(k)$ of ASD connections.}

Another type
of fixed points with $q\neq 0$,
which will be called  branch (ii),  are in the abelian
Coulomb phase ($D_\m \phi =0$ and $\phi\neq 0$).
In this branch the gauge symmetry is broken down to
$U(1)$ and  the vector bundle
$E$ reduces to the sum of line bundles
$E = \zeta \oplus \zeta^{-1}$ where
\eqn\swa{
\zeta\cdot \zeta = -k, \qquad
c_1(\zeta) =-c_1(\zeta^{-1}) \in H^2(X;\BZ).
}
The curvature two-form $F$
of $E$ reduces to
\eqn\swb{
F\rightarrow
\Fr{1}{2 i}
\left(
\matrix{{F_{3}} & 0\cr 0& -{F_{3}}}
\right)
\in \eufm{su(2)},
}
where $\Fr{1}{2}F_3$ is the curvature of the line bundle $\zeta$.
Now  equation \new\ becomes
\eqn\mcll{
 m q^{\dot\a}
+ i \phi_3 T_3 q^{\dot{\a}} = 0,\quad
\hbox{\it where}\quad \phi_a T^a\rightarrow
\phi_3 T_3=\Fr{1}{2i}\left(\matrix{\phi_3 & 0\cr 0 & -\phi_3\cr}\right),
}
which can be written as
\eqn\mclla{\eqalign{
m q_{(1)}^{\dot\a} + \Fr{1}{2}\phi_3 q^{\dot\a}_{(1)} = 0,\cr
m q_{(2)}^{\dot\a} -\Fr{1}{2}\phi_3 q^{\dot\a}_{(2)} = 0,\cr
}}
where
\eqn\fund{
q^{\dot\a} = \left(\matrix{q^{\dot\a}_{(1)}\cr q^{\dot\a}_{(2)}\cr}\right),
}
and $(1)$, $(2)$ denote the color index.
Thus the only nontrivial
solutions for $q^{\dot\a}$ are either
\eqn\wwa{
q^{\dot\a}=\left(\matrix{q^{\dot\a}_{(1)}\cr 0}\right)
\quad\hbox{\it and}\quad  2m +\phi_3 =0,
}
or
\eqn\wwb{
q^{\dot\a}=\left(\matrix{0\cr q^{\dot\a}_{(2)}}\right)
\quad\hbox{\it and}\quad  2m - \phi_3 =0.
}

Now the $Q_m$ fixed point equation \mfixed\
(or the monopole equation \monopole) becomes
\eqn\swc{\eqalign{
F^{+3}_{\m\n} +\Fr{1}{2i}
\left(q^\dagger_{(1)}\bar\s_{\m\n} q_{(1)}
-q^\dagger_{(2)}\bar\s_{\m\n} q_{(2)}
\right)&=0,
\cr
\left(\matrix{
\s^{\m}\eufm{D}_{\m+}& 0\cr
0 & \s^{\m}\eufm{D}_{\m-}
}\right)
\left(\matrix{q_{(1)}\cr
q_{(2)}}\right)&=0, \cr
}
}
where we have  either $q_{(1)}=0$ or $q_{(2)} =0$.
Here,  $\s^\m \eufm{D}_{\m\pm}$
denotes the abelian $spin^c$
the Dirac operator acting on $\G(W_\eufm{c}^+\otimes\zeta^{\pm1})$
satisfying the following  Weitzenb\"{o}k formula
\eqn\weizen{
(\s^\m \eufm{D}_{\m\pm})^2 = -g^{\m\n}\eufm{D}_{\m\pm} \eufm{D}_{\n\pm}
\mp F^{+}_{3\m\n} - p^+_{\m\n} + \Fr{1}{4}R,
}
and $p^+$ is the self-dual part of the curvature of
$det(W^+_\eufm{c})=L_\eufm{c}$.
If we exchange $A^3 \rightarrow -A^{3}$ and $q_{(1)} \rightarrow q_{(2)}$
the equation is symmetric.
Since we always have a pair of line  bundles $\pm\zeta$ for each
bundle reduction, we can always fix   $q_{(2)} = 0$ and  $q_{(1)} =M\neq 0$,
and regard the two solutions  as the same  equation
for the line bundles satisfying $\zeta\cdot \zeta = -k$.
We also set
$\s^\m\eufm{D}_{\m +}$ to $\s^\m \eufm{D}_\m$.

Now we have the celebrated Seiberg-Witten monopole
equations
\eqn\swe{\eqalign{
F^{+}_{3\m\n } + \Fr{1}{2 i}
M^\dagger\bar\s_{\m\n} M
=0,
\cr
\s^{\m} \eufm{D}_\m M
=0.
}}
Note that curvature $F^{+}_{3\m\n}$ is the
curvature of the line bundle $\zeta^2$,
while $M$ is the section of $W^+_\eufm{c}\otimes \zeta $.
Since $\zeta$ is an integral class, one can regard
$W^+_\eufm{c}\otimes \zeta = W^+_\eufm{c^\pr}$ as
a different $spin^c$ bundle for the different
$spin^c$ structure $\eufm{c^\pr}=\eufm{c}+2\zeta$,
i.e., $det(W^+_\eufm{c}\otimes \zeta) = L_\eufm{c}\otimes\zeta^2$.
The Weitzenb\"{o}ck formula \weizen,
which can be written as
\eqn\weizenb{
(\s^\m \eufm{D}_{\m})^2 = -g^{\m\n}\eufm{D}_{\m} \eufm{D}_{\n}
-(F^{+}_{3\m\n} + p^+_{\m\n}) + \Fr{1}{4}R,
}
also shows that $\s^\m \eufm{D}_{\m}$ is the $spin^c$ Dirac
operator acting on $W^+_\eufm{c^\pr}$. We will frequently
use the notation $\Fs{\eufm{D}}^{\eufm{c}^\pr}$ for
$\s^\m\eufm{D}_\m$.
The original  Seiberg-Witten equation consists of the
curvature of $det(W^+_\eufm{c^\pr})$ and the section of
 $W^+_\eufm{c^\pr}$ \WittenC.
The equation \swe\ should be viewed as a perturbed
Seiberg-Witten monopole equation for the
$spin^c$ structure $\eufm{c}^\pr$;
\eqn\sweb{\eqalign{
F^{\eufm{c}^\pr +}_{\m\n } +\Fr{1}{2 i}
M^\dagger\bar\s_{\m\n} M
=p^+_{\m\n},
\cr
\Fs{\eufm{D}}^{\eufm{c}^\pr} M
=0,
}}
where $F^{\eufm{c}^\pr} = F^{+}_{3\m\n} + p^+_{\m\n}$
denotes the curvature of $det(W^+_\eufm{c^\pr})$.

All this can also be seen from the action \act\ or \action.
The relevant part is the Cartan subalgebra part
of Eq.~\bosonic.
Note that
\eqn\swf{\eqalign{
 \Fr{1}{h^2}\int\!d^4\!x\sqrt{g}\biggl[
 &\Fr{1}{4}F^{+\m\n}_3 F^{+}_{3\m\n}
- \Fr{1}{2}p^+_{\m\n}M^\dagger \s^{\m\n}M
+  \Fr{1}{2}g^{\m\n}\eufm{D}_\m M^\dagger_{\dot\a}
    \eufm{D}_\n M^{\dot\a}
\cr
&-  \Fr{1}{16}(M^\dagger\bar\s^{\m\n}M )(M^\dagger\bar\s_{\m\n}M)
+ \Fr{1}{8}R (M^\dagger_{\dot\a s}M^{\dot\a })
\biggr],
}
}
which can be rewritten as
$
\Fr{1}{h^2}\!\int\!\!\!
\sqrt{g} d^4x\left(\Fr{1}{4}|s_3|^2 +\Fr{1}{2}|k_3|^2\right)
$, where $s_3 = F^+_{3\m\n} +\Fr{1}{2i}M^\dagger\bar\s^{\m\n}M$
and $k_3 = \s^\m \eufm{D}_\m M$. The perturbation can
be removed by replacing $F^+_{3\m\n}$ with
$F^+_{3\m\n} + p^+_{\m\n}$ in $s_3$. Then, we have
\eqn\swh{\eqalign{
\Fr{1}{h^2}\!\int\!\!\!\sqrt{g} d^4\!x
&\left(\Fr{1}{4}|s_3|^2 +\Fr{1}{2}|k_3|^2\right)
\cr
&=\Fr{1}{h^2}\int\!d^4\!x\sqrt{g}\biggl[
\Fr{1}{4}{F}^{\eufm{c}^\pr +\m\n}_3 {F}^{\eufm{c}^\pr +}_{\m\n}
-\Fr{1}{16}(M^\dagger\bar\s^{\m\n}M )
                  (M^\dagger\bar\s_{\m\n}M)
\cr
&\qquad\qquad\qquad
+\Fr{1}{2}g^{\m\n}\eufm{D}_\m M^\dagger_{\dot\a}
\eufm{D}_\n M^{\dot\a}
+\Fr{1}{8}R (M^\dagger_{\dot\a}M^{\dot\a})
\biggl].
}
}
That is, one can rewrite \swf\ in terms of
the the curvature of $det(W^+_\eufm{c^\pr})$, which
is equivalent to absorbing the term
$-\Fr{1}{2}p^+_{\m\n}M^\dagger\s^{\m\n}M$ by
a field redefinition.

Thus we have the localization to the moduli space
$\CM(\eufm{c}^\pr)$ of Seiberg-Witten monopole
with the $spin^c$ structure $\eufm{c}^\pr$, i.e.,
the space of solutions of \sweb\ in
$\CA_{det(W^+_\eufm{c^\pr})}\times \G(W^+_{\eufm{c}^\pr})$
modulo the gauge symmetry $S^1$.
The formal dimension of the moduli space $\CM(\eufm{c}^\pr)$  is
given by
\eqn\dimsw{
dim \CM(\eufm{c}^\pr)
=2 index \left(\Fs{\eufm{D}}^{\eufm{c}^\pr}\right)
-\Fr{\chi +\s}{2}
=
\Fr{ \eufm{c}^\pr\cdot\eufm{c}^\pr}{4} -\Fr{2\chi +3\s}{4}.
}
We will denote  a $spin^c$
structure $\eufm{c}^\pr$ by $x$ if $dim \CM(\eufm{c}^\pr)=0$,
i.e.,\foot{
This condition can be written by $\eufm{c}^\pr = \eufm{c} +2\zeta$
with
$$
\zeta\cdot\zeta < 0, \qquad \eufm{c}\cdot\zeta =
-index\left( \Fs{D}^E_\eufm{c}\right)
+ 2\D,
$$
where $D =(\chi +\s)/4$.
}
\eqn\secl{
x\cdot x = \Fr{2\chi +3\s}{4}.
}
Then the moduli space $\CM(x)$ consists of a finite collection
of points. The Seiberg-Witten invariant $n_x$ is the
algebraic sum of the number of  points counted
with sign.

Applying the fixed point theorem of Witten
for the global supersymmetry (see Sect.~3.1 of \WittenD),
the path integral can be written as the
sum of contributions of the branch (i) and the branch (ii).
So the path integrals can be written as a
certain sum of the Donaldson and the
Seiberg-Witten invariants.

Note that we should have chosen a $spin^c$ structure $\eufm{c}$
to define the twisting of the hypermultiplet.
And the TQCD depends on the choice of the $spin^c$ structure.
Now,
in branch (ii), one can view
the choice of different twisting as the choice of different
perturbation of the Seiberg-Witten equations.\foot{
Note that the perturbed term $p^+_{\m\n}$ in \sweb\
is the self-dual part of the curvature of $det(W^+_\eufm{c})$.
See \WittenC\
for the applications of the perturbation.}
It is shown that  the Seiberg-Witten invariant $n_x$
is independent of the (generic) perturbation \WittenC.
{\it Consequently, the family of TQCD parametrized by
the different choice of the $spin^c$ structure is governed
by the the same Seiberg-Witten invariants in  branch (ii)}.
This is a crucial property since we will use the path integral
of TQCD, which depends on the choice of  the $spin^c$ structure,
to obtain the path integral of  the TYM theory.

Before moving to the next topic,  we review the orientation of
both moduli spaces $\CM(k)$ and $\CM(x)$.
The proof of the orientability of a moduli space amounts
to showing the triviality of a determinant line  of
elliptic operator arising from the linearization of the
moduli space \DK. For the moduli space $\CM(k)$ of
ASD connection, the elliptic operator is $(d_A^+\oplus d_A^*)$.
Donaldson showed that an orientation of the space
$H^1(X;\BR)\oplus H^+(X;\BR)$ induces orientations of $\CM(k)$.
For the moduli space $\CM(x)$ of the Seiberg-Witten monopoles
the elliptic operator is $((d+d^*)\oplus \Fs{\eufm{D}}^x)$
and the triviality
of its determinant line was shown \WittenC.
The orientation of determinant line
of $(d+d^*)$ is fixed once and for all by
picking an orientation of  $H^1(X;\BR)\oplus H^+(X;\BR)$.
Since the determinant line bundle of Dirac operator  $\Fs{\eufm{D}}^x$
is naturally trivial, one can define an orientation of $\CM(x)$.
If we replace $x$ with $-x$ which corresponds to different
trivialization of the determinant line, we have
\eqn\nxmx{
n_{-x} = (-1)^\D n_x,\qquad \D=\Fr{\chi+\s}{4} = index\;\eufm{\Fs{D}}^x.
}

Since we will compare the contribution from the moduli space
$\CM(k)$ with those from the moduli spaces $\CM(x)$,
the relative orientations are important.
Since the orientations of $det\;ind(d_A^+\oplus d_A^*)$ and
$det\;ind(d+d^*)$ are governed by the same data
$H^1(X;\BR)\oplus H^+(X;\BR)$, the ambiguity in
the comparison can only come from the determinant line
of the Dirac operator $\eufm{\Fs{D}}^x$.
We will fix the orientation of $det\; ind(d_A^+\oplus d_A^*)$
to the opposite of $det\; ind(d+d^*)$.

\subsec{The Stationary Phases}

In this subsection, we will
address the problem (2) mentioned in  Sect.~$3.1$.
One can find the following combination
is $Q_m$ invariant
\eqn\mce{
\Fr{1}{2\pi}\int_X d^4\!x \sqrt{g}\left(
 m q^\dagger_{\dot\a} q^{\dot\a}
+i\phi^a q^\dagger_{\dot\a} T_a q^{\dot\a}
+\bar\p^{\dot{\a}}_{\tilde{q}}\bar\p_{q\dot\a}
\right),
}
which is $Q_m$ exact;
\eqn\mch{\eqalign{
&-i\left\{Q_m, -\Fr{1}{4\pi}\int_X\! d^4x \sqrt{g}\left(
q^\dagger_{\dot\a}\bar\psi^{\dot\a}_{\!\tilde q}
+\bar\psi_{\!q\dot\a}q^{\dot\a}\right)\right\}
\cr
&\qquad\qquad\qquad\qquad=
\Fr{1}{2\pi}\int_X d^4\!x \sqrt{g}\left(
 m q^\dagger_{\dot\a} q^{\dot\a}
+i\phi^a q^\dagger_{\dot\a} T_a q^{\dot\a}
+\bar\p^{\dot{\a}}_{\tilde{q}}\bar\p_{q\dot\a}
\right).
}
}
If we remove the term proportional to $m$ in the supersymmetry
transformation laws, the above term is no-longer
invariant under the original supercharge $Q$.
Instead, the $Q$ invariant combination is
\eqn\mcf{
\Fr{1}{2\pi}\int_X d^4x \sqrt{g}\left(
i\phi^a q^\dagger_{\dot\a} T_a q^{\dot\a}
+\bar\p^{\dot{\a}}_{\tilde{q}}\bar\p_{q\dot\a}
\right).
}

We add the $Q_m$-exact  term \mce\ to the  topological
action $S^\pr$ to get a one-parameter family of the topological theory
\eqn\mci{
S^\pr(t) = S^\pr +\Fr{t}{2\pi} \int_X\! d^4x \sqrt{g}\left(
m q^\dagger_{\dot\a} q^{\dot\a}
+i\phi^a q^\dagger_{\dot\a} T_a q^{\dot\a}
+\bar\p^{\dot{\a}}_{\tilde{q}}\bar\p_{q\dot\a}
\right).
}
If, in particular, we choose $t = 2\pi m/h^2$ we superficially
recover the full mass terms of  the hypermultiplet
as the physical theory.
Since the $t$ dependent term is $Q_m$-exact,
the theory does not depend on $t$ as long as $t\neq 0$ by the
standard argument of the cohomological theory.

Now we consider the topological correlation function
\eqn\mcj{
\left< {\hat v}^r\hat u^s \right>_{TQCD(m,t)}
= \Fr{1}{\hbox{vol}(\CG)}\int \CD Y
e^{-S^\pr - \Fr{t}{2\pi}\int_X\! d^4x \sqrt{g}\left(
m q^\dagger_{\dot\a} q^{\dot\a}
+i\phi^a q^\dagger_{\dot\a} T_a q^{\dot\a}
+\bar\p^{\dot{\a}}_{\tilde{q}}\bar\p_{q\dot\a}
\right) }\cdot {\hat v}^r \hat u^s.
}
In the $t=0$ limit, the above formula is identical to \mbcr.
Thus, the path integral \mcj\ can be evaluated in a suitable limit.
One can consider $t$ as purely imaginary
and take the limit
$Im(t) =  \infty$,
one may use the method of the stationary phases.
In the $Im(t)\rightarrow \infty$ limit, the dominant
contribution to the path
integral comes from the stationary phases (the critical points).
Such an approximation is exact, provided that we sum over
the contributions of the all critical points.

The equation for  the stationary phases  in the $t$-dependent terms
in \mcj\  is
\eqn\mcl{
 m q^{\dot\a}
+ i \phi_a T^a q^{\dot{\a}} = 0,\qquad
m q^\dagger_{\dot\a}+i q_{\dot\a}^{\dagger}\phi^{a}T_{a}=0.
}
Thus the stationary phase equation is identical
to the $Q_m$-fixed point equation,
$\d_m \bar\p^{\dot\a}_{\!\tilde{q}}
=\d_m \bar\p_{\!{q}\dot\a}=0$.
This is not surprising as can be seen from relation \mch.

Thus, the stationary phases have the same two
branches as discussed in the previous subsection.
Combining with the exact semi-classical limit
$h^2\rightarrow 0$, the exactness of the stationary
phase approximation recovers the same localization
of the path integral as predicted by the fixed point
argument of the global supersymmetry $Q_m$.

Consequently, one can use either the $Q_m$-fixed
point theorem or the combination of the
semi-classical and the stationary phase approximation.
Both methods say that one should evaluate the
path integral exactly at the two branches
and calculate Gaussian integral over the quadratic
terms due to the transverse degrees of freedom.

Before moving to the next topic we should remark
on three subtle points related to the stationary phase.
There are some criteria for the independence
on  a BRST trivial deformation \WittenE.

First of all,
there should be no new fixed points flowing
from infinity. In our case, the $t$-dependent
term does not change any fixed points of
the global supercharge $Q_m$. It is a rather
natural deformation for the theory having
the bare mass term.

Secondly, the additional BRST trivial term should preserve
the $U$-number symmetry of the original theory.
Note that the expression \mce\  contains the term with
$U$-number $0$ as well as the term with $U$-number $2$.
This means that the theory with $t\neq 0$ and
$t=0$ can be actually different.

Thirdly, the deformation term should not change the property
that the energy-momentum tensor is a $Q_m$ commutator.
It can  easily be seen that the particular deformation does not
alter the property.
However,  the $Q_m$-exact term \mce\ does not
preserve the global scaling invariance of the theory.
Showing the global scaling invariance amounts to
proving that the trace of energy-momentum tensor
is a total divergence. The failure of the property
can be seen by counting
the net scaling dimension  of the fields in \mce\
which  is $2$ rather than $4$.
We will return to this important  issue later on.

\subsec{The global $S^1$ symmetry}

In this subsection,
we study the global $U(1)$ symmetry on the hypermultiplet.
We will show that there are two different types of the $S^1$ fixed points
which are identical to the two branches of the stationary phases.

The theory has a global $U(1)$ symmetry acting on the hypermultiplet;
\eqn\sym{
q \rightarrow e^{i\theta} q,
}
which leaves the action\foot{This $S^1$ symmetry
should be read in general as
$Q_h \rightarrow e^{i\theta} Q_h$ and
$\tilde Q_h \rightarrow \tilde Q_h e^{-i\theta} $
which obviously leaves the action invariant.}
as well as the fixed point equation of the
global supersymmetry invariant
\eqn\monopolea{\eqalign{
F^{+a}_{\m\n} + q^\dagger\bar\s_{\m\n}T^a q &=0 ,
\cr
\s^{\m} D_\m q &=0.
}}
This $U(1)$ action has two branches of the fixed points.

The obvious fixed point is when $q=0$ and all the other fields belonging
to the hypermultiplet vanish as well.
Then the monopole equation \monopolea\ becomes the standard
anti-self-duality equation. We will call this fixed point  branch (i).
There is another type of the fixed point.
Note that the path integral is defined over the space of fields
modulo the local gauge symmetry.
Thus the $S^1$ action can have another fixed
point if there are gauge transformations such that
\eqn\hitchin{
g(\theta) q = e^{i\theta} q,
\qquad
g(\theta)^{-1}d_A g(\theta) = d_A.
}
The situation is very similar to the self-duality equations of
Hitchin \Hitchin.\foot{In ref.~\VW, the similar symmetry was
considered  for the twisted $N=4$ super-Yang-Mills theory which
is a close cousin of TQCD as well as of the Hitchin equations.
Vafa and Witten showed and computed that the path integral of
the theory is the Euler character of the instanton moduli space
provided that certain vanishing theorems hold.
The vanishing theorem
amounts to the absence of the branch (ii) contributions.
In our case, we will express the invariant due to the branch (i)
in terms of the contributions of the branch (ii).
It will be interesting to see if the method we develop in this paper
can be applied  for  computing   the Euler character and
testing the $S$-duality in a general case.
An obvious starting point will be to study the model
by adding the mass term which breaks $N=4$  down to
the $N=2$ supersymmetry.
}
The first equation implies that if $q\neq 0$, then $g(\theta)$ is
not an identity for $\theta \neq 2n\pi$ where $n$ is an integer.
And, then, the second equation
implies that the connection $A$ is reducible and
that the $SU(2)$ bundle
reduces to the direct sum of line bundles, i.e., $E = \zeta\oplus\zeta^{-1}$
{}.
Then, $g(\theta)$ becomes
diagonalized. Since $q$ belongs to the fundamental representation,
$q$ must be either $q = (q_{(1)}, 0)^T$ or
$q=(0,q_{(2)})^T$ to satisfy the first equation.

Thus, two branches of the fixed point of $S^1$
action are identical to two branches of the stationary phases.

\subsec{The $\CG\times S^1$-equivariant Cohomology}

The relation between  the $Q_m$-fixed points and the fixed points of
the $S^1$ action suggests that $Q_m$ has a close relation with
the $S^1$-equivariant cohomology.
The commutation relation \commut\ implies that one can identify
$-im$ as the generator of the $Fun(Lie(S^1))$ associated with the
global $S^1$ action on the hypermultiplet.
If we consider the space
$\CA\times \G(W^+_\eufm{c}\otimes E)$,
the global $S^1$-symmetry
acts on $(A,q, q^\dagger)$ by $(A,e^{i \theta}q,
q^\dagger e^{-i \theta})$.
The $\CG$ action
on $\CA\times \G(W^+_\eufm{c}\otimes E)$
is generated by vector fields  $V(\phi)$.
We can consider the algebra  of polynomial functions
$\hbox{Fun($Lie(S^1)$)}$ generated by $-im$.
Now, one can define the (infinite dimensional)
$\CG\times S^1$-equivariant de Rham cohomology.
The equivariant de Rham complex is defined by
\eqn\maax{\eqalign{
\O^*_{\CG\times S^1}(\CA\times &\G(W^+_\eufm{c}
\otimes E)
\cr
&=
\left(\O^*(\CA\times \G(W^+_\eufm{c}\otimes E))
\otimes
Fun(Lie(\CG))\otimes Fun(Lie(S^1)) \right)^{\CG\times S^1}.
\cr
}
}
The associated differential operator is $\d$ which can be
formally represented as
\eqn\mabx{
\d = -\int  d^4\!x \sqrt{g}\P^I(x)\Fr{\d}{\d A^{I}(x)}
+i\int  d^4\!x \sqrt{g}V(\phi(x))^I\Fr{\d}{\d \P^{I}(x)}
+\int  d^4\!x \sqrt{g}V(m)^J\Fr{\d}{\d \P^{J}(x)}.
}
We have
\eqn\macx{
\d^2 = -\int  d^4\!x \sqrt{g}(\CL_{V(\phi(x))} +\CL_{V(m)}).
}
Thus, $\d^2 = 0$ on the $\CG\times S^1$-invariant subspace
$\O^*_{\CG\times S^1}(\CA\times \G(W^+_\eufm{c}
\otimes E)$ of
$\O^*(\CA\times \G(W^+_\eufm{c}\otimes E))
\otimes \hbox{Fun}(Lie(\CG))
\otimes \hbox{Fun}(Lie(S^1))$.
The $\CG\times S^1$-equivariant de Rham cohomology
$H^*_{\CG\times S^1}(\CA\times \G(W^+_\eufm{c}
\otimes E))$ is defined as
the pairs
$$
(\O^*_{\CG\times S^1}(\CA\times \G(W^+_\eufm{c}
\otimes E)),\;\d).
$$
The basic supersymmetry algebra
\eqn\madx{
\eqalign{
\hat\delta_m A_\m =& + i\vr \lambda_\m,\cr
\hat\delta_m q^{\dot\a}=
 &-\vr \bar{\psi}_{\!\tilde{q}}^{\dot{\alpha}},
  \cr
  \hat\delta_m q_{\a}^{\dagger}=
 &-\vr \bar{\psi}_{\!q \dot{\alpha}} ,
 \cr
}
\qquad\eqalign{
\hat\delta_m \lambda_\m =& -\vr D_\m \phi,\cr
\hat\delta_m \bar{\psi}_{\!\tilde{q}}^{\dot{\alpha}}=
& -i\vr \phi^{a}T_a q^{\dot{\a}} -\vr m q^{\dot\a},
  \cr\hat\delta_m \bar{\psi}_{\!q\dot{\alpha}}=
&+i\vr q_{\dot\a}^{\dagger}\phi^{a}T_{a}
 +\vr m q^\dagger_{\dot\a},
}\qquad\eqalign{
\hat\delta_m \phi = 0,\cr
\hat\delta_m m = 0,\cr
}}
shows that
the twisted supercharge of the theory with the massive hypermultiplet
can be interpreted
as the generator of the $\CG\times S^1$-equivariant de Rham cohomology
$H^*_{\CG\times S^1}(\CA\times \G(W^+_\eufm{c}
\otimes E))$.

Now, the problem of the $U$-number can be resolved.
We define the degree of the $\CG\times S^1$-equivariant complex
by the formula
\eqn\degr{
\hbox{deg}(\a\otimes\b\otimes\g) = \hbox{deg}(\a) +2\hbox{deg}(\b)
+2\hbox{deg}(\g),
}
for $\a \in \O^*(\CA\times \G(W^+_\eufm{c}
\otimes E))$,
$\b \in  \hbox{Fun}(Lie(\CG))$ and $\g \in \hbox{Fun}(Lie(S^1))$.
That is, we assign $U=2$ to the mass $m$,\foot{The interpretation
of the $S^1$-equivariant cohomology generator as a parameter
or vise versa is not new \AB.} regarding $m$ as an operator
or a constant field.
Then $Q_m$ increases the degree by one, as expected.
Note that the action $S^\pr$ of \acadd\ now has the $U$-number
zero. The expression
\mce\ also has the correct $U$-number $2$.
The familiar topological observable $\hat v$ \mba\
can be viewed as $\CG\times S^1$-equivariant extension
of  a differential two-form on
$\CA\times \G(W^+_\eufm{c} \otimes E)$.
Although $\hat v$ has the same form in the TYM, the massless
TQCD and the massive TQCD, due to the differences
between \remabc, \remab\ and  \mcdd,  it is effectively
different in each theory.

It is interesting to note that the last term in \mce\ is a closed
form. The first  and the second terms in \mce\ give the $S^1$ and
$\CG$-equivariant extensions of the last term, respectively.
By comparing \mce\ with \mch,
one can view the changing of the supersymmetry
from $Q$ to $Q_m$
as the recipe to introduce the term
\eqn\mcg{
\Fr{1}{2\pi}\int_X\!  d^4x \sqrt{g}
\left( m q^\dagger_{\dot\a} q^{\dot\a}\right),
}
in the supersymmetric way.
The TQCD with one hypermultiplet  depends on
the choice of the $spin^c$ structure $\eufm{c}$.
Thus, we have a family of TQCD parametrized
by the space of $spin^c$ structures on $X$.
Since the TQCD depends on a  choice of the $spin^c$ structure
$\eufm{c}$, it may be possible to embed the moduli space
$\CM(k)$ of anti-self-dual connections to the moduli space
$\CM(k,\eufm{c})$ as a connected component of
the fixed points locus of global $S^1$ action on
$\CM(k,\eufm{c})$ by varying the $spin^c$
structure.
The $S^1$ action has also another type of the fixed points
whose locus is the moduli space of the
(abelian) Seiberg-Witten
monopoles.
Thus, in the formal level,  the path integral evaluation
can be quite similar to the Duistermaat-Heckmann (DH)\foot{
If we consider the K\"{a}hler case,
the term  \mcg\ without $m$ can be identified with the momentum map
(Hamiltonian) of the $S^1$ action on
$\CA\times \G(W^+_\eufm{c}\otimes E)$
Thus, the analogy between  the DH integration formula and
our path integral computation becomes much closer
on K\"{a}hler manifolds.
This is one way to see why the massive deformation dramatically
enhances the computability of the path integral.}
integral formula \DH\ or the equivariant integration formula
of Berline and Vergne \BGV\AB.
However, such an interpretation is clearly problematic
unless the moduli space $\CM(\eufm{c},k)$ has highly
favorable properties.
We shall see that our path integral computation
gives a perfectly concrete formula in any case.

An alternative formal viewpoint is to consider the
equivariant $S^1$ localization from the beginning without
referring to the moduli space $\CM(\eufm{c},k)$.

 \subsec{A Synthesis}

The massless TQCD  has almost
the same properties and problems as the TYM theory.
The cohomological interpretation of those two theories
leads us to some integrals of differential forms
over the moduli spaces  which are  rarely compact.
Even though we assume the case when the moduli space
is actually compact, it is rarely possible to compute such an
integral explicitly.

Now the role of introducing the bare mass term to the hypermultiplet
becomes clear. The massive deformation we introduced
further localizes
the path integral to the moduli space $\CM(k)$ of ASD connections
and the moduli spaces $\CM(x)$ of the (abelian) Seiberg-Witten
monopoles. The latter spaces are compact and, if we assume
the simple type condition, they are zero-dimensional.
Although the path integral contributed from the moduli
space $\CM(k)$ would be  almost impossible to compute,
we can certainly get explicit results for the contributions
of $\CM(x)$.

We have seen the above localization
by the various arguments which are closely related
with each other.  It is quite amusing to see that
the massive deformation of TQCD leads to a beautiful
synthesis of the various aspects  of the
cohomological field theory and the equivariant
cohomology.   However, some subtle points
remain to be resolved.

The localization due to the global supersymmetry
is based on a very general assumption.
For example,  even if we regard the action $S^\pr$ \acadd\
as the complete action functional, the supersymmetry
transformation law \mcd\ predicts that the path integral
should be localized according to equation \new.
However, such a localization can be seen only after
adding the $Q_m$ exact term \mce\ to the action.
In TYM theory and the massless TQCD, one can recover
the same localization by the semi-classical limit as
predicted by the $Q$-fixed point arguments.
In fact, there is a drawback in the semi-classical
analysis \WittenA.
The kinetic energy term
for the scalar fields $\phi$ and $\bar \phi$ is not positive definite.
After twisting it seems to be more natural to regard those fields
as independent fields.
One can regard that $\phi$ is real
and $\bar\phi$ is purely imaginary.
Then, the localization $D_\m \phi =0$
can be seen by the stationary phase for $h^2\rightarrow 0$.
If we maintain the
complex conjugation relation $\phi \sim -\bar\phi^{*}$
for $\phi$ and $\bar\phi$ as
the physical theory, the kinetic energy is positive definite and the
localization can be seen by the usual semi-classical
limit.

The power of Witten's fixed point theorem is that
it does not refer to such complications.
However, this does not necessarily mean that
we don't need to add the remaining mass term \mce\
to the action. To ensure the correct localization
of the path integral, we should include all terms
that produce all the relevant fixed point equations of
the global supersymmetry by the equations of the
motion. For example, even the problematic kinetic term
for the scalar fields may not be necessary. However,
the fixed point theorem says that one
should replace a field by its fixed point value.
In the TYM theory or in the branch (i) of our theory,
the fixed point value for $\phi$ should be
zero to avoid reducible instantons.
Thus, the above replacement is obviously incorrect.
The resolution is to try to integrate out $\phi$ which results in
the  replacement of \remabc. The kinetic term for the scalar
fields is crucial.

In our viewpoint,  one should
maintain all the terms coming from the physical action.
In fact, we can readily justify this. In the physical theory,
the supercharge comes from the conserved supercurrent
which can be calculated by the various terms in the
action. The supersymmetry algebra then naturally follows.
That is, the particular supersymmetry responsible for
the crucial fixed point equation \new\ originates
from the full mass term for the hypermultiplet.
The twisting procedure just couples one of
the spinor indices to the internal global symmetry
of the theory.   Thus, we can not discard any term coming from
the physical theory.
Actually, our  computation
of the path integral will confirm the above assertion.

Now, the remaining problem is that the crucial term
\mce\ in the exponent of \mci\ violates the global
symmetries of the theory.
We argue that the mass term should carry the
$U$-number $2$.
This means that the theories with
$t=0$ and $t \neq 0$ are different, although
there are no new fixed points flowing from infinity,
since the positive ghost number of \mce\
will effect the ghost number anomaly cancellation
due to the observables.\foot{One may still add
the original term \mce\ to the action and control
the theory if one can.}
By choosing $t = 1/m$,
we can make the $t$-dependent term in \mci\
to carry the $U$-number $0$,
\eqn\deform{
\Fr{1}{2\pi m} \int_X\! d^4x \sqrt{g}\left( m q^\dagger_{\dot\a} q^{\dot\a}
+i\phi^a q^\dagger_{\dot\a} T_a q^{\dot\a}
+\bar\p^{\dot{\a}}_{\tilde{q}}\bar\p_{q\dot\a}
\right),
}
such that
the deformed theory does not depend on the additional
$Q_m$-exact term.
Now we can exponentiate the observable $\hat v$
\eqn\csb{
\left< \exp(\hat v+\tau \hat u)\right>_{m,\eufm{c}, k}
=\Fr{1}{\hbox{vol}(\CG)}\int \CD Y e^{-S^\pr -\Fr{1}{2\pi}
 \int_X\! d^4x \sqrt{g}\left(
q^\dagger_{\dot\a} q^{\dot\a}
+i\phi^a q^\dagger_{\dot\a} T_a q^{\dot\a}/{m}
+\bar\p^{\dot{\a}}_{\tilde{q}}\bar\p_{q\dot\a}/{m}
\right) +\hat v +\tau\hat u},
}
such that
\eqn\csc{
\left< \exp(\hat v +\tau \hat u)\right>_{m,\eufm{c}, k}
=\sum_{r +2s =d(\eufm{c},k)}   \Fr{(2\tau)^s}{r!s!}
\left< \hat v^{r}\hat u^{s}\right>_{m,\eufm{c}, k}.
}

The final problem is that
such a deformation  violates  the global scaling invariance.
If we scale the metric by a constant $g\rightarrow t g$ both $S^\pr$ and
$\hat v^\pr$ remain invariant. On the other hand
the deformation term \deform\  is scaled as $t^2$.
Fortunately, this property does not change the independence
of the correlation function on the metric.\foot{
The scaling independence
can be seen by showing that  the trace of
energy-momentum tensor is a total divergence, which
is independent of the $Q_m$-exactness of the energy-momentum
tensor. Eventually, we will set $m=0$. This limit
is smooth since the additional term can be
written as $\Fr{1}{m}\{Q_m, O\}$ and there is
no obstruction for going to $m=0$. Such an argument
is not valid in general, see the footnotes $(10)$ and $(11)$
of \WittenA.}
Thus, {\it we can  identify the stationary phase limit as
the large scaling limit of the metric}.
Actually, {\it only the large scaling limit of the metric
is the true stationary phase limit}.
One may introduce
an extra parameter and take the infinite limit. However,
its effect can be absorbed by a rescaling of the metric.
Since we are dealing with the metric invariant theory,
the large scaling limit of the metric is the true stationary
phase limit.  It is quite amusing to see that the localization
of the path integral, which is the key step of this paper,
is achieved by the large scale limit
of the metric.

In the physical theory, the analogous
step would be taking the infinite mass limit such that the hypermultiplet
can be integrated out.  Then,
the theory will reduce
to the theory without the matter field. Though this is the
most natural step to localize the theory to  branch (i),
it is not  clear how the path integral localizes to
branch (ii) as well. On the other hand, the physical
theory has the asymptotic freedom
so that it is scale-dependent.
The $N=2$ SYM theory interpolates
the two branches by the genuine quantum
scaling behaviours.
The seminal work of Seiberg
and Witten shows that the key simplification responsible
for the Seiberg-Witten invariants  occurs in the
strong coupling vacuum which is equivalent
to the weakly coupled vacua of massless $U(1)$
hypermultiplet.

In our case, the twisted theory with additional matter multiplet
having the bare mass in the large scale limit
of the metric localizes to the two branches
corresponding to two different limits of the physical
$N=2$ SYM theory.   Furthermore,  condition \wwa\
says that the $U(1)$ hypermultiplet is massless in
branch (ii).
{\it This is an amazing property.
How is it that the essentially classical treatment  of
a theory understands
the genuine quantum property of a different theory?}
The only answer to this question seems to be the
self-duality of the critical theory \SWb.
All those properties of
the asymptotically free theories and their twisted versions
can be some remnants of the critical theory through the
massive deformations.

In the remaining sections, we will concretely realize the
above picture. As all the three different viewpoints
of the localization suggest,  the path integral amounts to
evaluating  exactly at the locus of the each branch
and computes the Gaussian integrals of quadratic terms
due to the transverse degree.
The favorable interpretation is to take the large scaling limit first.
We define the action $S_m$  by
\eqn\actionm{\eqalign{
S_m = &S^\pr + \Fr{1}{2\pi} \int_X\! d^4x \sqrt{g}\left(
q^\dagger_{\dot\a} q^{\dot\a}
+i\phi^a q^\dagger_{\dot\a} T_a q^{\dot\a}/{m}
+\bar\p^{\dot{\a}}_{\tilde{q}}\bar\p_{q\dot\a}/{m}
\right)
\cr
= &S
+\Fr{1}{h^2}\int_X d^4 x \sqrt{g}
\left(2 im \bar\phi^a q^\dagger_{\dot\a} T_a q^{\dot\a}
+m\psi^{\a}_{\!\tilde{q}}\psi_{\!q\a}\right)
\cr
&+ \Fr{1}{2\pi} \int_X\! d^4x \sqrt{g}\left(
q^\dagger_{\dot\a} q^{\dot\a}
+i\phi^a q^\dagger_{\dot\a} T_a q^{\dot\a}/{m}
+\bar\p^{\dot{\a}}_{\tilde{q}}\bar\p_{q\dot\a}/{m}
\right).
}}
where we use the action $S$ in the form of Eq.~\act.

Picking a Riemann metric $g$, we rescale $g\rightarrow t g$
and take $t\rightarrow \infty$ limit.
In  branch (i), the gauge symmetry is unbroken
and the matter fields decouple as the transverse degrees of
freedom. The dominant contribution to the path
integral comes from the path integral of the TYM theory.
In  branch (ii), the gauge symmetry breaks down to
$U(1)$ and the hypermultiplet reduces to one of its
color. The suppressed color degrees of freedom for hypermultiplet
and the components of the $N=2$ vector multiplet
which do not belong to the Cartan subalgebra part
become the transverse degrees of freedom.
In the infinite scaling limit, it is sufficient to keep
only quadratic terms for the transverse degrees
and compute the one-loop approximations which are
arbitrarily good.

On the other hand, the path integrals
for  the non-transverse degrees should be computed
exactly. These path integrals correspond to the
path integral of TYM theory in  branch (i) and
the path integral of topological QED (Seiberg-Witten theory)
in branch (ii).

We will use the notations
$\left< O\right>_{m,\eufm{c},k}$, $\left< O\right>_{\eufm{c},k}$,
$\left< O\right>_{k}$, and $ \left< O\right>$ for the
correlation functions evaluated in the massive TQCD for given
$spin^c$ structure and instanton number,
in the massless TQCD for given
$spin^c$ structure and instanton number,
in the TYM theory for given
instanton number  and in the TYM theory with summation over
all instanton numbers, respectively.

\newsec{The Computation of the Path Integral}

In this section, we compute the path integral in the
large scaling limit of the metric.

\subsec{Branch (i) and the Donaldson-Witten theory}

In this branch, the degree of freedom for the hypermultiplet
becomes transverse.
One can decompose the action $S_m$ into
two parts
\eqn\pppt{
S_m \approx S_m(i) + \d^{(2)}S_m(i).
}
where  $ \d^{(2)}S_m(i)$ denotes  the quadratic
action due to
the transverse degrees (the matter fields $Q_h$ and
$\tilde Q_h$).
The action $S_m(i)$ in  branch (i) locus  reduces to
the familiar action of the TYM theory \WittenA
\eqn\tym{\eqalign{
S_m(i)=
\Fr{1}{h^2}\int\!&\sqrt{g}d^4\!x\biggl[
\Fr{1}{4}F^{+\m\n}_a F^{+a}_{\m\n}
-\Fr{1}{2}g^{\m\n}(D_\m\bar\phi)_a(D_\n\phi)^a
+\Fr{1}{8}[\phi,\bar\phi]_a[\phi,\bar\phi]^a
-i\chi^{\m\n}_a[\phi,\chi_{\m\n}]^a
\cr
&
+\chi^{\m\n}_a(d_A\l)^{+a}_{\m\n}
+\Fr{i}{2}g^{\m\n}(D_\m\eta)_a\l^a_\n
-\Fr{i}{2}g^{\m\n}[\l_\m,\bar\phi]_a \l_\n^a
+\Fr{i}{8}[\phi,\eta]_a \eta^a,
\cr
}
}
which is the standard action for the TYM theory.
Or, equivalently
\eqn\mmma{
\eqalign{
S_m(i)
= &-i\{Q_m,V_T(i)\}
\cr
=& -i\left\{Q, \Fr{1}{h^2}\int\sqrt{g} d^4\!x\biggl[
   \chi^{\m\n}_a
      \left(
       H^a_{\m\n} - i F^{+a}_{\m\n}
      \right)
- \Fr{1}{2}g^{\m\n}(D_\m\bar\phi)_a \l^a_{\n}
+ \Fr{1}{8}[\phi,\bar\phi]_a \eta^a
\biggl]
\right\},
\cr
}
}
and integrate $H_{\m\n}$ out. Note that $Q=Q_m$.

The contribution of the branch (i) to the correlation
function $\left< e^{\hat v}\right>_{\eufm{c},m,k}$
can be written as
\eqn\yoa{
\left< e^{\hat v}\right>_{\eufm{c},m,k}(i)
=\Fr{1}{\hbox{vol}(\CG)} \int \CD W e^{ - S(i) + \hat v(i)}
\times
\int\CD \tilde Q_h \CD Q_h  e^{-\d^{(2)}S_m(i) + \d^{(2)}\hat v(i)},
}
where $\hat v(i)$ denotes the usual observable for
TYM theory, $(\CD W)$ denotes the path integral
measure for TYM theory
and $\d^{(2)}\hat v(i)$ denotes the quadratic
term of $\hat v$ due to  the transverse degrees.
Note that  $\d^{(2)}\hat v(i)=0$.

\subsec{Transverse Path Integral for the Branch (i) }

The quadratic action due to the matter fields is
\eqn\actm{\eqalign{
\d^{(2)}S_m(i) =\Fr{1}{h^2}\int\! &d^4\!x\sqrt{g}\biggl[
-2 X_{\!\tilde{q}}^{\a} X_{\!q\a}
+{i} X_{\!\tilde{q}}^{\a}\s^{\m}{}_{\a\dot\a}D_\m q^{\dot\a}
+{i}D_\m q^\dagger_{\dot\a}\bar\s^{\m\dot\a\a}X_{\!q\a}
+\Fr{h^2}{2\pi}q^\dagger_{\dot\a} q^{\dot\a}
\cr
&
-{i}D_\m\bar\psi_{\!q\dot\a}\bar\s^{\m\dot\a\a}\psi_{\!q\a}
-{i}\psi_{\!\tilde{q}}^\a\s^{\m}{}_{\a\dot\a}
 D_\m\bar\psi_{\!\tilde{q}}^{\dot\a}
+2m\psi^\a_{\!\tilde{q}}\psi_{\! q\a}
+\Fr{h^2}{2\pi m}\bar\p^{\dot{\a}}_{\tilde{q}}\bar\p_{q\dot\a}
\biggr].\cr
}
}
The Gaussian integrals over auxiliary fields
$\psi^\a_{\!\tilde{q}},\psi_{\! q\a}, X_{\!\tilde{q}}^{\a}$ and $ X_{\!q\a}$
give
\eqn\actn{\eqalign{
\d^{(2)}S_m(i) =\Fr{1}{h^2}\int\! & d^4\!x\sqrt{g}\biggl[
\Fr{1}{2} D_\m q^\dagger_{\dot\a}\bar\s^{\m\dot\a\a}
     \s^{\n}{}_{\a\dot\a}D_\n q^{\dot\a}
+\Fr{h^2}{2\pi}q^\dagger_{\dot\a} q^{\dot\a}
\cr
&
-\Fr{1}{2m}D_\m\bar\psi_{\!q\dot\a}\bar\s^{\m\dot\a\a}
      \s^{\n}{}_{\a\dot\a}D_\n\bar\psi_{\!\tilde{q}}^{\dot\a}
-\Fr{h^2}{2\pi m}\bar\p_{q\dot\a}\bar\p^{\dot{\a}}_{\tilde{q}}
\biggr],
}
}
with the following determinant;
\eqn\bbt{
\Fr{
\biggl[det\left(2m\right)\biggr]_{(\p_{\!q}^{\a},\p_{\!\tilde{q}\a})}
}{
\biggl[det\left(-\Fr{1}{\pi}\right)\biggr]_{(X^{\a}_{\!\tilde{q}},X_{\!q\a})}
}
=
\biggl[det\left(-2\pi m\right)\biggr]_{\G(W^-_\eufm{c}\otimes E)}.
}
Usually the infinite dimensional determinants are not well
defined and need regularization. However, the above
determinant ratio is perfectly well defined due to the
global supersymmetry.
Now we are left with
\eqn\zza{
\int \CD q^\dagger\CD q\CD \bar\p_{\tilde{q}}\CD \bar\p_{q}
e^{
-\Fr{1}{h^2}\int\! d^4\!x \sqrt{g} \left(
\Fr{1}{2}q^\dagger\left[\Fs{D}^2 +\Fr{1}{\pi} h^2\right] q
+\Fr{1}{2m}\bar\p_{q}\left[\Fs{D}^2 +\Fr{1}{2} h^2\right]
 \bar\p_{\tilde{q}}
\right)}.
}
If we perform the
Gaussian integral over $(q^\dagger_{\dot\a}, q^{\dot\a})$
and $(\bar\psi_{\!q\dot\a}, \bar\psi_{\!\tilde{q}}^{\dot\a})$
by ignoring $\d^{(2)}\hat v(i)$
term, we have the following determinant\foot{
If we apply the Weitzenb\'{o}ck formula, $\Fs{D}^2$
contains the usual connection Laplacian, the scalar curvature
of metric and the curvature for connection. If we scale
$g \rightarrow t g$, all the terms scale out as $t^{-2}$.
The following formula clearly shows that the determinant
ratio  is independent of such a scaling.}
\eqn\bbtb{
\Fr{
\biggl[det\left( -\Fr{\Fs{D}^2 +\Fr{1}{2\pi} h^2}{m}
\right)\biggr]_{(\bar\p_{\!q\dot\a},\bar\p^{\dot\a}_{\!\tilde{q}})}
}{
\biggl[det\left(\Fr{1}{2\pi}(\Fs{D}^2 +\Fr{1}{2\pi} h^2)
\right)\biggr]_{(q^\dagger_{\dot\a},q^{\dot\a})}
}
=
\biggl[det\left(-\Fr{2\pi}{m}\right)\biggr]_{\G(W^+_\eufm{c}\otimes E)}.
}
Combining \bbt\ and \bbtb, we have
\eqn\zzf{
\int \CD Q_h \CD \tilde Q_h e^{-\d^{(2)}S_m(i)}
=\biggl[
det\left(-\Fr{2\pi}{m}\right)
\biggr]_{\G(W^+_\eufm{c}\otimes E) \ominus \G(W^-_\eufm{c}\otimes E)}
= \left(-\Fr{2\pi}{m}\right)^{index(\Fs{D}_\eufm{c}^E)}.
}
Thus, we have
\eqn\uuha{
\left< e^{\hat v}\right>_{\eufm{c},m,k}(i)
=\left(-\Fr{2\pi}{m}\right)^{index(\Fs{D}_\eufm{c}^E)}
\times \left< e^{\hat v}\right>_k.
}

\subsec{Branch (ii) and the Seiberg-Witten Theory}

In this branch, the gauge symmetry breaks down to
$U(1)$ (the maximal torus). The components of any
field which do not belong to the Cartan subalgebra part
becomes the transverse variable.
One can decompose the action $S_m$ into
two parts
\eqn\pppt{
S_m \approx S_m(ii) + \d^{(2)}S_m(ii).
}
To begin with, we regard that action $S_m(ii)$
consists of the Cartan subalgebra part of all the adjoint
fields
\eqn\actionii{\eqalign{
S_m(ii)=
\Fr{1}{h^2}&\int\!d^4\!x\sqrt{g}\biggl[
\Fr{1}{4}F_3^{+\m\n} F^{+}_{3\m\n}
+\Fr{i}{4}p^{+}_{\m\n}(M^{\dot\a\dagger}\bar\s^{\m\n}M^{\dot\a})
+\Fr{1}{2}g^{\m\n}\eufm{D}_\m M^\dagger_{\dot\a} \eufm{D}_\n M^{\dot\a}
\cr
&
- \Fr{1}{16}(M^\dagger\bar\s^{\m\n} M )
                  (M^\dagger\bar\s_{\m\n} M)
+\Fr{1}{8}R (M^\dagger_{\dot\a}M^{\dot\a})
-\Fr{1}{2}g^{\m\n}\rd_\m\bar\phi_3 \rd_\n\phi_3
\phantom{\biggr]}
\cr
&
+\chi^{\m\n}_3(d\l_3)^{+}_{\m\n}
-\Fr{1}{2}\chi^{\m\n}_3\bar\psi_{\!M}\bar\s_{\m\n} M
+\Fr{1}{2}\chi^{\m\n}_3 M^\dagger\bar\s_{\m\n}\bar\psi_{\!\tilde{M}}
+\Fr{i}{2}g^{\m\n}\rd_\m\eta_3 \l_{3\n}
\phantom{\biggr]}
\cr
&
+\Fr{1}{2i} M^\dagger_{\dot\a}\eta_3  \bar\psi_{\!\tilde M}^{\dot\a}
-\Fr{1}{2i} \bar\psi_{\!M\dot\a}\eta_3  M^{\dot\a}
+\Fr{1}{2i}M^\dagger_{\dot\a}\l_{\m}  \bar\s^{\m\dot\a\a}\psi_{\!M\a}
+\Fr{1}{2i}\psi_{\!\tilde{M}}^\a \s^{\m}{}_{\a\dot\a}\l_{3\m}
M^{\dot\a}
\cr
&
-{i}\eufm{D}_\m\bar\psi_{\!M\dot\a}\bar\s^{\m\dot\a\a}\psi_{\!M\a}
-{i}\psi_{\!\tilde{M}}^\a\s^{\m}{}_{\a\dot\a}
 \eufm{D}_\m\bar\psi_{\!\tilde{M}}^{\dot\a}
- \bar\psi_{\!M\dot\a}\bar\phi_3  \bar\psi_{\!\tilde{M}}^{\dot\a}
+\psi_{\!\tilde{M}}^\a  (\phi_3 + 2m)\psi_{\!M\a}
\phantom{\biggr]}
\cr
&
+\Fr{1}{2} M^\dagger_{\dot\a} M^{\dot\a}\bar\phi_3(\phi_3 +2m)
+\Fr{h^2}{2\pi}(1 +\Fr{\phi_3}{2m})M^\dagger_{\dot\a}M^{\dot\a}
+\Fr{h^2}{2\pi m}\bar\psi_{\!M\dot\a} \bar\psi_{\!\tilde{M}}^{\dot\a}
\biggr].\cr
}
}
The above action has the following supersymmetry
\eqn\susya{\eqalign{
\hat\delta_m A_{3\m} &= i\vr \lambda_{3\m},\cr
\hat\delta_m \lambda_{3\m} &= -\vr \rd_\m \phi_3,\cr
\hat\delta_m \phi_3 &= 0,\cr
}\qquad
\eqalign{
\hat\delta_m \chi_{3\m\n}
&= \Fr{i}{2}\vr(F^+_{3\m\n} +\Fr{1}{2i}M^\dagger\bar\s_{\m\n}M),\cr
\hat\delta_m \bar\phi_3 &= i\vr\eta_3,\cr
}\qquad\eqalign{
\hat\delta_m H_{3\m\n} &=0,\cr
\hat\delta_m \eta_3 &=0,\cr
}
}
and
\eqn\susyb{
\eqalign{
\hat\delta_m M^{\dot\a}
=&-\vr \bar{\p}_{\!\tilde{M}}^{\dot{\alpha}},
\qquad\qquad
\hat\delta_m \bar{\psi}_{\!\tilde{M}}^{\dot{\alpha}}
= -\Fr{1}{2}\vr(\phi_3 + 2m)M^{\dot\a},
 \cr
\hat\delta_m M_{\a}^{\dagger}
=&-\vr \bar{\psi}_{\!M \dot{\alpha}} ,
\qquad\quad
\hat\delta_m \bar{\psi}_{\!M\dot{\alpha}}
=+\Fr{1}{2}M^\dagger_{\dot\a}\vr(\phi_3 + 2m),
 \cr
\hat\delta_m \psi_{\!M\a}
=& - \Fr{i}{2}\vr \s^{\m}{}_{\a\dot\a}
\eufm{D}_\m M^{\dot\a} ,
\cr
\hat\delta_m \psi_{\!\tilde M}^{\a}
=&+ \Fr{i}{2}\vr \eufm{D}_\m M_{\dot\a}^\dagger
      \bar \s^{\m\dot\a\a}.
}
}
One can view the reduced action $S_m(ii)$ as
a purely abelian theory which is the twisted version
of the $N=2$ super-Maxwell theory coupled with
hypermultiplet having the bare mass.
The fixed point equations
$\hat\delta_m \chi_{\m\n} =\hat\delta_m \psi_{\!M\a} =0$
show that this theory describes the Seiberg-Witten invariants.
Another important fixed point equation
$\hat\delta_m \bar{\psi}_{\!\tilde{M}}^{\dot{\alpha}}=0$
shows that
one can replace $\phi$ with $-2m$,
\eqn\csl{
\phi \rightarrow \phi_3 T_3 = \Fr{1}{2 i}
\left( \matrix{\phi_3&0\cr 0&-\phi_3}\right)
= \Fr{1}{i}\left(\matrix{-m&0\cr 0&m}\right).
}
Of course, the above theory should be viewed as
a subsector of the bigger theory. More precisely,
we should interpret the theory as an effective theory
in one of the two types of branches in the large
scaling limit.
As an embedded theory, the fixed
point equation $(\phi + 2m) = 0$ implies that the
$U(1)$ hypermultiplet is massless.
Then our previous discussion clearly shows that  branch
(ii) of massive QCD in the large scale limit is described
by the $U(1)$ massless hypermultiplet.
The condition  $(\phi + 2m) = 0$ also implies
that the $U(1)$ hypermultiplet becomes massless
at the $S^1$-fixed points.
Up to now, we have regarded the bare mass
$m$ as a constant field carrying the $U$-number $2$.
This interpretation was based on the identification
of $m$ with the $S^1$-equivariant cohomology generator.
Since this branch is in the fixed point of $S^1$ action
we can treat $m$ as a number.
{\it This is because branch (ii) are $S^1$-fixed points
where we undo the $S^1$ action by the local $U(1)$
gauge symmetry}. This can also be seen
by writing $S_m(ii)$ honestly in the large scaling limit.
Then the terms involving $m$ in \actionii\ decouple
from the path integral.
Of course, $m$ should be
treated  as carrying the $U$-number $2$ in the transverse
integrations.

The contribution of  branch (ii) to the correlation
function $\left< e^{\hat v}\right>_{\eufm{c},m,k}$
can be written as
\eqn\yooa{
\eqalign{
\left< e^{\hat v}\right>_{\eufm{c},m,k}(ii)
=&\Fr{1}{\hbox{vol}(\CG)}
\int \CD \tilde Q_{h(1)} \CD  Q_{h(1)} \CD W_3
\int \CD W e^{ - S(ii) + \hat v(ii)}
\cr
&\times
\int \CD \tilde Q_{h(2)} \CD Q_{h(2)} \CD W_+\CD W_-
e^{-\d^{(2)}S_m(ii) + \d^{(2)}\hat v(ii)}.
}
}

The contribution from
\eqn\uiou{
\Fr{1}{\hbox{vol}(\CG_3)}
\int\CD \tilde Q_{h(1)} \CD Q_{h(1)}  \CD W_3
e^{ - S(ii) + \hat v(ii)},
}
can be easily determined. First of all,
in a simple type manifold we only need to consider
the zero-dimensional moduli space of the Seiberg-Witten
monopoles. Then there are no fermionic zero-modes.
So one can simply replace $\hat v(ii)$ with its
fixed point values
\eqn\repps{
\hat v(ii) = \Fr{m}{2\pi}(v\cdot x).
}
One can expand the action around the fixed points (a point in
the zero dimensional moduli space $\CM(x)$) up to the quadratic
term.
The action has the following general form:
\eqn\actffr{
 \Fr{1}{h^2}\int\!d^4\!x\sqrt{g}\biggl[
\Phi\Delta_B\Phi+i\Psi D_F\Psi\biggr],
}
where $\Phi$ and $\Psi$ denote the bosonic and
fermionic fields, respectively, and $\Delta_B$
and $D_F$ are their corresponding operators.
The similar situation  is discussed, in detail, in \WittenA.
The Gaussian integral gives
\eqn\dratio{
\Fr{\hbox{Pfaff} D_F}{\sqrt{det \Delta_B}} =\pm 1.
}
The sign depends on a choice of orientation.
To make the assignment of the sign  meaningful, one
should prove  that the determinant
line bundle of $D_F$ (called the Pfaffian line bundle)
can be trivialized.  Note that  $D_F$ is precisely
the linearization of the abelian Seiberg-Witten
equations. Thus the orientatibility  of  Pfaffian line bundle
follows from the orientability of the moduli space
$\CM(x)$.
Thus we can read the result
immediately as
\eqn\raaaf{
\left< 1\right>_x = \CN n_x,\qquad\hbox{ where } \qquad
n_x =\sum_{s \in\CM(x)}(-1)^{\e_s},
}
where $\CN$ denotes the standard renormalization
due to the local operators constructed from
metric depending only on $\chi$ and $\s$ \WittenA\WittenB\VW.
Consequently, we have
\eqn\uiou{
\int \CD \tilde Q_{h(1)} \CD Q_{h(1)} \CD W_3
\int \CD X e^{ - S(ii) + \hat v(ii)}
=\CN \sum_{x} n_x e^{ \Fr{m}{2\pi}(v\cdot x)}.
}

Before leaving this subsection, we should address the
problem of the compatibility between the
replacement $\phi\rightarrow -2m$ and
$\phi \rightarrow \left<\phi\right>$. In the TYM theory,
there should be no zero-modes of $\phi$ (the non-zero
solution of $\phi$ for $D_\m \phi = 0$).
Since all the topological observables contain $\phi$,
the replacement of $\phi$ with its zero modes
gives the vanishing results. The correct field theoretical
treatment is to replace $\phi$ with its expectation value
$\left<\phi\right>$ which amounts to integrating it out \WittenB.
In the massive TQCD, $\phi$ has the zero-modes in
branch (ii) and the preferred value is $-2m$.
This can be seen from the fixed point equation \susyb\
as well as the large scaling limit, which we are considering,
for  the deformation term
\eqn\csh{
\Fr{1}{2\pi} \int_X\! d^4\!x \sqrt{g}\left(
q^\dagger_{\dot\a} q^{\dot\a}
+i q^\dagger_{\dot\a}\Fr{\phi^a T^a}{m} q^{\dot\a}
+\Fr{\bar\p^{\dot{\a}}_{\tilde{q}}\bar\p_{q\dot\a}}{m}
\right).
}
At  branch (ii)  the above
expression is zero by definition.
Of course,  this is equivalent to the substitution \csl.
This means that the expectation value $\left<\phi\right>$
reduces to $\phi(ii)$ in \csl\ for the effective $U(1)$ theory.
The question is how this can be consistent with the
substitution due to \mcdd.

The relevant part is  the $\phi_3$
component
in \mcdd, we have
\eqn\cte{\eqalign{
-\left(q^\dagger_{\dot{\a}}q^{\dot\a}\right)\left<\phi_3\right>(ii)
=&-\left(D_\m D^\m\right)\left<\phi_3\right>(ii) -i[\l^\m,\l^\n]_3
\cr
&
+ 2 \bar\psi_{\!q\dot{\a}}\left(\matrix{1 & 0\cr 0 &-1}\right)
\bar\psi^{\dot\a}_{\!\tilde q} + 2 m q^\dagger_{\dot\a}
\left(\matrix{1 & 0\cr 0 &-1}\right) q^{\dot\a}.
}}
In the branch (ii) stationary phase or the $Q_m$-fixed point,
$\phi_3$ is a non-zero covariant constant and there are no-zero
modes of $\bar\psi_{\!q\dot{\a}}$, $\bar\psi^{\dot\a}_{\!\tilde q}$
and $\l_\m$. Furthermore, we have $\bar\psi_{\!q\dot{\a}(2)}
=\bar\psi^{\dot\a(2)}_{\!\tilde q}
=q^\dagger_{\dot\a(2)} =q^{\dot\a(2)}=0$.
Thus, the above formula reduces to
\eqn\ctg{
\left(q^\dagger_{\dot{\a}(1)}q^{\dot\a(1)}
\right)\left<\phi_3\right>(ii)
= - 2 m q^\dagger_{\dot\a(1)} q^{\dot\a(1)}.
}
The  solution for $\left<\phi_3\right>(ii)$ is $-2m$ as desired.

The formula \cte\ is useful to extract the precise form
of the quadratic terms
for
$i q^\dagger_{\dot\a}\phi^a T^a q^{\dot\a}$
due to the transverse variables
$\bar\psi_{\!q\dot{\a}(2)}, \bar\psi^{\dot\a(2)}_{\!\tilde q},
q^\dagger_{\dot\a(2)}, q^{\dot\a(2)}$.
{}From \cte, we have the relation
\eqn\cth{
-\Fr{1}{2}\left( q^\dagger_{\dot{\a}(2)}q^{\dot\a(2)}
\right)\left<\phi_3\right>(ii)
= \Fr{i}{2}\left<\phi_3\right>(ii) [A_\m, A^{\m}]_3-\Fr{i}{2}[\l^\m,\l_\m]_3
-\bar\psi_{\!q\dot{\a}(2)}
\bar\psi^{\dot\a}_{\!\tilde q(2)}
-m q^\dagger_{\dot\a(2)} q^{\dot\a(2)}.
}
Thus,  the quadratic term for the deformation
term \csh\ is
\eqn\cti{\eqalign{
\delta^{(2)}\biggl(\Fr{1}{2\pi} \int_X\! d^4\!x \sqrt{g}&\left(
q^\dagger_{\dot\a} q^{\dot\a}
+i q^\dagger_{\dot\a}\Fr{\phi^a T^a}{m} q^{\dot\a}
+\Fr{\bar\p^{\dot{\a}}_{\tilde{q}}\bar\p_{q\dot\a}}{m}
\right)\biggr)(ii)
\cr
&= \int_X\! d^4\!x \sqrt{g}\left(
\Fr{1}{\pi}A^+_\m A^{\m-}
+\Fr{\l^+_\m\l^{\m -}}{2\pi m}
\right).
}
}
Note that this quadratic expansion is compatible with
the obvious choice
\eqn\cticom{
\Fr{1}{2\pi}\int_X\! d^4\!x \sqrt{g}
\left( q^\dagger_{\dot\a(2)} q^{\dot\a(2)}
+\Fr{\bar\p^{\dot{\a}}_{\tilde{q}(2)}\bar\p_{q\dot\a(2)}}{m}\right),
}
due to the $Q_m$-fixed point equation\foot{Obviously, the
quadratic expansion in the
neighborhood of the fixed point locus should be taken
in the $Q_m$ invariant way.},
$
\hat\delta_m \chi_{\m\n}
= F^{+a}_{\m\n} +q^\dagger\bar\s_{\m\n}T^a q =0
$.

\subsec{Transverse  Path Integral for  Branch (ii)}

In this branch the gauge symmetry is broken  down to $U(1)$
and the $\pm$ components of the adjoint fields\foot{
Note that  $W$ decomposes as
$
W=W_3 T_3 + W_+T_+ + W_-T_-.
$
}
and the
components of the hypermultiplet with the suppressed color
index  become the transverse variable.
Basically, we will integrate out all the transverse degrees.
The relevant quadratic action is given by
\eqn\qactionii{
\eqalign{
\d^{(2)} S_m(ii)
=
 \Fr{1}{h^2}
 &\int\!
d^4\!x\sqrt{g}\biggl[
4H^{\m\n}_+ H_{\m\n -}
-2i H^{\m\n}_+ (\eufm{D}A_-)^+_{\m\n}
-2i H^{\m\n}_- (\eufm{D}A_+)^+_{\m\n}
+4\w\chi^{\m\n}_+\chi_{\m\n-}
\cr
&+2\chi^{\m\n}_+(\eufm{D}\l_-)^+_{\m\n}
+2\chi^{\m\n}_-(\eufm{D}\l^+)_{\m\n}
-2 X_{\!\tilde{q}}^{\a (2)} X_{\!q\a}^{(2)}
+i X_{\!\tilde{q}}^{\a (2)}\s^{\m}{}_{\a\dot\a} \eufm{D}_\m q^{\dot\a (2)}
\phantom{\biggr]}
\cr
&+i \eufm{D}_\m q^{\dagger (2)}_{\dot\a}\bar\s^{\m\dot\a\a}X_{\!q\a}^{(2)}
-\w\psi_{\!\tilde{q}}^{\a (2)} \psi_{\!q\a}^{(2)}
-i\psi_{\!\tilde{q}}^{\a (2)}\s^{\m}{}_{\a\dot\a}
\eufm{D}_\m\bar\psi_{\!\tilde{q}}^{\dot\a (2)}
\phantom{\biggr]}
\cr
&-i\eufm{D}_\m\bar\psi_{\! q}^{\dot\a (2)}
\bar\s^{\m\dot\a\a}\psi_{\!q\a}^{(2)}
+\Fr{1}{4}\w^2\bar\phi_+\bar\phi_-
-i\w \bar\phi_+(\eufm{D}_\mu A^{\mu}_-)
+i \w\bar\phi_-(\eufm{D}_\mu A^{\mu}_+)
\phantom{\biggr]}
\cr
&
+\Fr{1}{4}\w\eta_+ \eta_-
+i\eta_+(\eufm{D}_\mu\l^\mu_-)
+i\eta_-(\eufm{D}_\mu\l^\mu_+)
+\Fr{h^2}{\pi} A_{\m+} A^{\m}_{-}
+\Fr{h^2}{2\pi m}\l_{\m+}\l^{\m}_{-}
\biggr],\cr
}
}
where $\pm$ in the subscript denotes the $T_\pm$ components
while $+$ in the superscript denotes the self-dual part,
and the substitution $\w = -2m$ is understood

Now we evaluate the transverse integral
\eqn\trarrr{
\Fr{1}{\hbox{vol}(\CG_\pm)}
\int \CD \tilde Q_{h(2)} \CD Q_{h(2)} \CD W_+\CD W_-
e^{-\d^{(2)}S_m(ii) + \d^{(2)}\hat v(ii)}.
}
Note the ordering of the path integral measure
should be used consistently.
We choose unitary gauge in which
\eqn\gauge{
\phi_{\pm}=0,
}
where
\eqn\decom{
\phi=\w T_3 + \phi_+T_+ + \phi_-T_-.
}
In this gauge $\phi$ has values on the maximal torus
(Cartan sub-algebra).
By following the standard Faddev-Povov gauge fixing,
we introduce a new
nilpotent BRST operator
$\d$ with the algebra
\eqn\gaugea{\eqalign{
\d \phi_{\pm} = \pm i c_{\pm} \w, \qquad \d c_{\pm}=0,
\cr
\d \phi_3  =0, \qquad \d \bar{c}_{\pm}=b_{\pm},  \qquad \d b_{\pm}=0,
\cr
}}
where $c_{\pm}$ and $\bar{c}_{\pm}$ are anti-commuting ghosts
and anti-ghosts,
respectively, and
$b_{\pm}$ are commuting auxiliary fields.
The action for gauge fixing terms reads
\eqn\agau{\eqalign{
S_{m,gauge}(ii) =
  &\d\biggl[\Fr{1}{h^2}\int_X i\bigl(\bar{c}_-*\w_+  + \bar{c}_+*\w_-
\bigr)\biggr]
 \cr
= &
\Fr{1}{h^2}\int_X \biggl[i\bigl(b_-*\phi_+  + b_+*\phi_- \bigr)
-\bar{c}_{-}*(\w)c_{+} +\bar{c}_{+}(\w)c_{-}\biggr].
 \cr
}}
The integrations over the auxiliary fields $b_\pm$ lead
to the gauge fixing condition \gauge.
The Gaussian
integrations over $\bar{c}$
and $c$ give
\eqn\deta{
\biggl[det\left(\w\right)\biggr]^{1/2}_{(c_+, \bar c_-)}
\biggl[det\left(\w\right)\biggr]^{1/2}_{(\bar c_+, c_-)}.
}

Now consider the transverse part involving
$\bar\phi_\pm$ and $\eta_\pm$.
The quadratic action relevant to this sector is given by
\eqn\acta{
\eqalign{
\d^{(2)}S_m(ii)= \Fr{1}{h^2}\int\!
&d^4\!x\sqrt{g}\biggl[
\Fr{1}{4}\w^2\bar\phi_+\bar\phi_-
-i\w \bar\phi_+(\eufm{D}_\mu A^{\mu}_-)
+i \w\bar\phi_-(\eufm{D}_\mu A^{\mu}_+)
\cr
&
+\Fr{1}{4}\w\eta_+ \eta_-
+i\eta_+(\eufm{D}_\mu\l^\mu_-)
+i\eta_-(\eufm{D}_\mu\l^\mu_+)
\biggr].\cr
}
}
The integrations over $(\bar\phi_+,\bar\phi_-)$ and
over $(\eta_+,\eta_-)$ combined
with \deta\ give
\eqn\detb{\eqalign{
\biggl[det\left(\Fr{\w^2}{4\pi}\right)\biggr]^{-1/2}_{(\phi_+,\phi_-)}
\biggl[det\left(\Fr{\w}{4}\right)\biggr]^{1/2}_{(\eta_+,\eta_-)}
\biggl[det\left(\w\right)\biggr]^{1/2}_{(c_+, \bar c_-)}
&\biggl[det\left(\w\right)\biggr]^{1/2}_{(\bar c_+, c_-)}
\cr
&\equiv \biggl[det(-2\pi m)\biggr]^{1/2}_{\O^0(\pm)}.
}
}
The transverse part involving
$H^{\m\n}_\pm$ and  $\chi^{\m\n}_\pm$
is given by
\eqn\actb{
\eqalign{
\d^{(2)}S(ii)= \Fr{1}{h^2}\int\!
&d^4\!x\sqrt{g}\biggl[
4H^{\m\n}_+ H_{\m\n-}
-2i H^{\m\n}_+ (\eufm{D}A_-)^+_{\m\n}
-2i H^{\m\n}_- (\eufm{D}A_+)^+_{\m\n}
\cr
&
+4\w\chi^{\m\n}_+\chi_{\m\n-}
+2\chi^{\m\n}_+(\eufm{D}\l_-)^+_{\m\n}
+2\chi^{\m\n}_-(\eufm{D}\l_+)^+_{\m\n}\biggr].
\cr
}
}
The integrations over $(H_+, H_-)$ and over $(\chi_+,\chi_-)$
give
\eqn\detc{
\biggl[det(4\w)\biggr]^{1/2}_{(\chi_+,\chi_-)}
\times \biggl[det\left(\Fr{4}{\pi}\right)\biggr]^{-1/2}_{(H_+,H_-)}
\equiv \biggl[det(-2\pi m)\biggr]^{1/2}_{\O^{2+}(\pm)}.
}
The transverse part involving
$X_{\!\tilde{q}}^{\a (2)}, X_{q\a}^{(2)},
\psi_{\!q\a}^{(2)}$ and $\psi_{\!\tilde{q}}^{\a (2)}$ is
\eqn\actc{
\eqalign{
\d^{(2)}S(ii)= \Fr{1}{h^2}\int\!
&d^4\!x\sqrt{g}\biggl[
-2 X_{\!\tilde{q}}^{\a (2)} X_{\!q\a}^{(2)}
-\w\psi_{\!\tilde{q}}^{\a (2)} \psi_{\!q\a}^{(2)}
\cr
&
+i X_{\!\tilde{q}}^{\a (2)}\s^{\m}{}_{\a\dot\a} \eufm{D}_\m q^{\dot\a (2)}
+i \eufm{D}_\m q^{\dagger (2)}_{\dot\a}\bar\s^{\m\dot\a\a}X_{\!q\a}^{(2)}
\cr
&-i\psi_{\!\tilde{q}}^{\a (2)}\s^{\m}{}_{\a\dot\a}
\eufm{D}_\m\bar\psi_{\!\tilde{q}}^{\dot\a (2)}
-i\eufm{D}_\m\bar\psi_{\! q}^{\dot\a (2)}
\bar\s^{\m\dot\a\a}\psi_{\!q\a}^{(2)}
\biggr].
\cr
}
}
The integrations over $(X_{\!\tilde{q}}^{\a (2)}, X_{q\a}^{(2)})$
and over $(\psi_{\!q\a}^{(2)}, \psi_{\!\tilde{q}}^{\a (2)})$
give
\eqn\detd{
\biggl[det\left(-\Fr{1}{\pi}\right)\biggr]^{-1}_{
(X_{\!\tilde{q}}^{\a (2)}, X_{q\a}^{(2)})}
\times
\biggl[det(2m)\biggr]^{1/2}_{(\p_{\!q\a}^{(2)},\p_{\!\tilde{q}}^{\a (2)}) }.
}

Now we collect all the remaining terms
which came from the various Gaussian integrations
\eqn\actd{\eqalign{
\d^{(2)}S_m(ii)=
\Fr{1}{h^2}\int\!d^4\!x\sqrt{g}\biggl[
(\eufm{D}_\m A^{\m}_-)(\eufm{D}_\n A^{\n}_+)
-\Fr{1}{4}(\eufm{D} A_-)^{+\m\n} (\eufm{D} A_-)^{+}_{\m\n}
+\Fr{h^2}{\pi} A_{\m-} A^{\m}_{+}
\cr
+\Fr{1}{2m}\biggl(
(\eufm{D}_\m \l^{\m}_-)(\eufm{D}_\n \l^{\n}_+)
-\Fr{1}{4}(\eufm{D}\l_-)^{+\m\n} (\eufm{D}\l_+)^{+}_{\m\n}
 +\Fr{h^2}{\pi}\l_{\m+}\l^{\m}_{-}\biggr)
\cr
+\Fr{1}{2}\Fs{\eufm{D}} q^{\dagger(2)}_{\dot\a}
\Fs{\eufm{D}} q^{\dot\a (2)}
-\Fr{1}{2m}\Fs{\eufm{D}} \bar\psi_{\!q\dot\a}^{(2)}
\Fs{\eufm{D}}\bar\psi_{\!\tilde{q}}^{\dot\a (2)}
\biggr].
}}
The Gaussian integrals over $(q^{\dagger (2)}_{\dot\a},q^{\dot\a(2)})$
and over $(\bar\psi_{\!q\dot\a}^{(2)}, \bar\psi_{\!\tilde{q}}^{\dot\a (2)})$
give
\eqn\dete{
\biggl[det\left(\Fr{\Fs{\eufm{D}}^2}{4\pi}\right)\biggr]^{-1}_{
(q^{\dagger (2)}_{\dot\a}, q^{\dot\a(2)})}
\biggl[det\left(-\Fr{\Fs{\eufm{D}}^2}{2m}\right)\biggr]_{
(\bar\psi_{\!q\dot\a}^{(2)},\bar\psi_{\!\tilde{q}}^{\dot\a (2)})}.
}

Now we are only left with $A_\pm$ and $\l_\pm$
whose quadratic action can be rewritten
in a compact form
\eqn\effective{
\d^{(2)}S(ii)
=
\Fr{1}{h^2}\int_X
\biggl(
A_+\wedge *(\nabla+\Fr{h^2}{\pi}) A_{-}
-\Fr{1}{2m}\l_+\wedge *(\nabla +\Fr{h^2}{\pi})\l_{-}
\biggr)  ,
}
where
$\nabla \equiv \eufm{D}\eufm{D}^* -\Fr{1}{4}\eufm{D}^{+*}\eufm{D}^{+}$.

If we integrate over
$(A_+, A_-)$ and over $(\l_+,\l_-)$, we get
\eqn\detf{
\biggl[det\left(\Fr{\nabla +h^2/\pi}{2\pi} \right)\biggr]^{-1/2}_{(A_+,A_-)}
\times
\biggl[det\left(-\Fr{\nabla +h^2/\pi}{m} \right)\biggr]^{1/2}_{(\l_+,\l_-)}
\equiv
\biggl[det(-2\pi m)\biggr]^{-1/2}_{\O^1(\pm)}  .
}
Collecting all  the determinant ratios  \detb, \detc, \detd, \dete\
and \detf, we have
\eqn\detg{\eqalign{
\biggl[det(-2\pi m)&\biggr]^{1/2}_{\Omega^0(\pm)
\ominus\Omega^1(\pm)\oplus\Omega^{2+}(\pm)}
\biggl[det(-2\pi m)\biggr]^{-1}_{[\bar\psi^{\dot\a (2)}_{\!\tilde{q}}]
\ominus[\psi_{\!q\a}^{(2)}]}
\cr
=&(-2\pi m)^{-\Fr{1}{2}\times index_\pm (\eufm{D}^+ + \eufm{D}^*)}
(-2\pi m)^{-(index(\Fs{D}_\eufm{c}^E) -\Fr{1}{8}({x}\cdot{x} -\s))}
\cr
=&(-2\pi m)^{-\Fr{1}{2}\times index(d_A^+ + d_A^*) -\D}
\cdot(-2\pi m)^{-index(\Fs{D}_\eufm{c}^E) +\D}
\cr
=&
(-2\pi m)^{-(4k-3\D)}\cdot(-2\pi m)^{-index (\Fs{D}^E_\eufm{c})}
\cr
=&
(-2\pi m)^{ - d(k) - index (\Fs{D}^E_\eufm{c})}
\cr
=& (-2\pi m)^{ - \Fr{1}{2}\times dim \CM(\eufm{c},k)}
=(-2\pi m)^{-d(\eufm{c},k)}.
\cr
}
}
In the above, we used
\eqn\fact{\eqalign{
\Fr{1}{2}\times index\left(d_A^+ + d_A^*\right)
\equiv d(k)
&= \Fr{1}{2}\left(dim H_A^1 - dim H_A^0 - dim H^2_A\right)
\cr
&= 4k -dim(G)\D,
}}
where $H_A^i$ denotes the three cohomology groups
of the instanton complex (See for example \DK).\foot{
Note that $dim (SU(2)) =3$. Since we already fixed gauge,
we have
$index (\eufm{D}^+ + \eufm{D}^*) =index(d_A^+ + d_A^*)$.
The action of $\eufm{D}^+ + \eufm{D}^*$ to the gauge
singlet (the Cartan subalgebra part) is identical to
that of $(d^+ + d^*)$ which contributes $-2\D$ to the
$index (\eufm{D}^+ + \eufm{D}^*)$. So we have
$$
index_\pm (\eufm{D}^+ + \eufm{D}^*)
= index(d_A^+ + d_A^*) + 2\D.
$$
We used a similar procedure for the Dirac index.
}
We also used
\eqn\factc{
index \Fs{D}_\eufm{c}^E = - k + \Fr{rank(E)}{8}(\eufm{c}\cdot\eufm{c} -\s),
}
and
\eqn\factb{
\D = \Fr{1}{8}({x}\cdot{x} -\s),
}
which follows from the zero-dimensionality of the abelian
Seiberg-Witten monopole moduli space $\CM(x)$.
As a check, we consider the case when $dim \CM(x) = 2 n$
such that
\eqn\checka{
n =  \Fr{1}{8}({x}\cdot{x} -\s) - \D.
}
The formula \detg\ becomes
\eqn\checkb{
(-2\pi m)^{-d(k) - index\Fs{D}_\eufm{c}^E + n}
= (-2\pi m)^{-(dim \CM(\eufm{c},k) - dim \CM(x))/2},
}
which is consistent.

Finally we evaluate the path integral
\eqn\final{\eqalign{
\int \CD A^+\CD A^- \CD \l^+\CD\l^-
&e^{-
\Fr{1}{h^2}\int_X
\biggl(
A_+\wedge *(\nabla+\Fr{h^2}{\pi}) A_{-}
+\Fr{1}{2m}\l_+\wedge *(\nabla +\Fr{h^2}{\pi})\l_{-}
\biggr)
}
\cr
&\qquad \times\exp\left(\Fr{1}{4\pi^2}\int_\S\left(
m A_{+}\wedge A_{-}
+ \Fr{1}{2}\l_{+}\wedge \l_{-}
\right)\right).
}
}

The path integral \final\ can easily be done using the elementary
techniques  of quantum field theory. (See Appendix A.)
The first step is to determine the  Green's functions.
For $A_+$ and $A_-$,
we have
\eqn\zzb{
\D_F (x_1 -x_2)
= \int \Fr{d^4p}{(2\pi)^4}\sqrt{g}
e^{-ip\cdot (x_1-x_2)}\Fr{h^2}{p^2 + \Fr{h^2}{\pi}},
}
where $p$ denotes the Fourier transformed
variable or the four-momentum such that
\eqn\zzc{
\left(\nabla + \Fr{h^2}{\pi}\right)
\D_F (x_1 -x_2) =h^2 \d^{(4)}(x_1-x_2).
}
The integral \zzb\ is obviously divergent in the
ultraviolet (the large momentum).
However, the infinite  scaling limit of the metric in
a compact manifold
is identical to the infinitesimally  small limit of the momentum.\foot{
It  may be tempting to think that a similar thing would happen
if we set $h^2\rightarrow 0$, which is not the case.
The key simplification of the theory comes from the
large scaling limit of the metric rather than the
semi-classical limit.
 }
Thus, the above integral is simply a delta function;
\eqn\rra{
\D_F(x_1-x_2) = \pi \d^{(4)}(x_1-x_2).
}
The similar analysis for
$\l_+$ and $\l_-$ shows their propagator is given
by
\eqn\rrb{
\D_F(x_1-x_2) = 2\pi m\d^{(4)}(x_1,x_2).
}
Now we can  perform the Gaussian integral which
gives
\eqn\rrc{
\exp\left({\Fr{m^2}{4\pi^2}\left(\Fr{v\cdot v}{2}\right)}\right),
}
together with the determinant \detf.
Thus, we find
\eqn\thusw{\eqalign{
\Fr{1}{\hbox{vol}(\CG_\pm)}
\int \CD \tilde Q_{h(2)} \CD Q_{h(2)}&  \CD W_+\CD W_-
e^{-\d^{(2)}S_m(ii) + \d^{(2)}\hat v(ii)}
\cr
&=
\left(-\Fr{2\pi}{m}\right)^{d(k) + index(\Fs{D}_\eufm{c}^E)}
e^{\Fr{m^2}{4\pi^2}\left(\Fr{v\cdot v}{2}\right)}.
}
}

Combining \thusw\ with \uiou, we have the total
contribution of  branch (ii),
\eqn\fiallii{
\left< e^{\hat v}\right>_{m,\eufm{c},k}(ii)
=\CN\left(-\Fr{2\pi}{m}\right)^{d(k) + index(\Fs{D}_\eufm{c}^E)}
\sum_x n_x e^{\Fr{m^2}{4\pi^2}\left(\Fr{v\cdot v}{2}\right)+
\Fr{m}{2\pi}(v\cdot x)}.
}

\subsec{The Results}

Collecting everything  in the previous subsections,
we get
\eqn\csq{\eqalign{
&\left< e^{\hat v}\right>_{m,\eufm{c},k}
\cr
&\quad=
-\left(-\Fr{2\pi}{m}\right)^{d_0(\eufm{c},k)}
\left< e^{\hat v} \right>_{k}
+\CN\left(-\Fr{2\pi}{m}\right)^{d(\eufm{c},k)}
\sum_x n_x
e^{\Fr{m}{2\pi} (v\cdot x)+\Fr{m^2}{4\pi^2}\left(\Fr{v\cdot v}{2}\right)},
\cr
&\quad=
-(-1)^{d_0(\eufm{c},k)}\left(\Fr{2\pi}{m}\right)^{d_0(\eufm{c},k)}
\biggl[\left< e^{\hat v} \right>_{k}
-(-1)^{\D}\CN\left(\Fr{2\pi}{m}\right)^{d(k)}
\sum_x n_x
e^{\Fr{m}{2\pi} (v\cdot x)+\Fr{m^2}{4\pi^2}\left(\Fr{v\cdot v}{2}\right)}
\biggr],
}
}
where $d_0(\eufm{c},k) =index\left(\Fs{D}_\eufm{c}^E\right)$,
and $d(\eufm{c},k) = d_0(\eufm{c},k)+d(k)$.
We multiplied  the factor $-1$ to $\left< e^{\hat v} \right>_{k}$
since we introduced the opposite orientations for
$det(d_A^+\oplus d_A^*)$  relative to $det(d+d^*)$.
Note that we have an additional relative sign $(-1)^\D$
between the contributions of the two branches.
If we replace $m$ with $-m$, we have
\eqn\csqa{\eqalign{
&\left< e^{\hat v}\right>_{-m,\eufm{c},k}
\cr
&\quad =
-\left(\Fr{2\pi}{m}\right)^{d_0(\eufm{c},k)}
\biggl[
\left< e^{\hat v} \right>_{k}
-\CN\left(\Fr{2\pi}{m}\right)^{d(k)}
\sum_x n_x
e^{-\Fr{m}{2\pi} (v\cdot x)+\Fr{m^2}{4\pi^2}\left(\Fr{v\cdot v}{2}\right)}
\biggr]
\cr
&\quad=
-\left(\Fr{2\pi}{m}\right)^{d_0(\eufm{c},k)}
\biggl[
\left< e^{\hat v} \right>_{k}
-(-1)^{\D}\CN\left(\Fr{2\pi}{m}\right)^{d(k)}
\sum_x n_x
e^{\Fr{m}{2\pi} (v\cdot x)+\Fr{m^2}{4\pi^2}\left(\Fr{v\cdot v}{2}\right)}
\biggr],
}
}
where we have used
\eqn\nyui{\eqalign{
\sum_x n_x
e^{-\Fr{m}{2\pi} (v\cdot x)+\Fr{m^2}{4\pi^2}\left(\Fr{v\cdot v}{2}\right)}
&=
(-1)^\D \sum_x n_{-x}
e^{\Fr{m}{2\pi} (v\cdot(- x))+\Fr{m^2}{4\pi^2}\left(\Fr{v\cdot v}{2}\right)}
 \cr
&=
(-1)^\D\sum_x n_x
e^{\Fr{m}{2\pi} (v\cdot x)+\Fr{m^2}{4\pi^2}
\left(\Fr{v\cdot v}{2}\right)}.
}
}
Thus the relative sign $-(-1)^\D$ between the two
branches remains unchanged. This is a crucial test
since the theory without the bare mass or without
hypermultiplet should be independent of $m$.
Note also that the relative sign  $-(-1)^\D$ does not
depend on the $spin^c$ structure chosen to
define TQCD. This is also an important test of consistency
since the result for the theory without hypermultiplet (TYM)
should be independent of whatever TQCD we are using.
We would like to emphasize again, as  explained
in Sect.~$3.2$, that the Seiberg-Witten
invariants in \csq\ or in \csqa\ are independent of
the $spin^c$ structure which defines a particular
TQCD.

This is a judicious moment to determine all the  invariants.
Since the left-hand side of \csq\ or \csqa\ is regular for $m=0$,
the non-$zero$th powers in $m$ of
the right-hand side should vanish order by order.
In particular
\eqn\csr{
\left<\exp({\hat v})\right>_{k}
= (-1)^\D\CN \sum_{r=0}^{[d(k)/2]}\Fr{ (v\cdot v)^r}{2^r(d(k)-2r)!r!}
\sum_{x}n_x(v\cdot x)^{d(k)-2r}.
}
This is a universal relation independent of
the family of TQCD parametrized by the space of the
$spin^c$ structure.
At this stage, we can determine the normalization term
$\CN$ by comparing with the known results, which
turns out to be
\eqn\crss{
\CN = (-1)^\D 2^{2+\Fr{1}{4}\left(7\chi + 11\s\right)},
}
where the last power of $2$ appears in similar
fashion with  \WittenC\WittenB. The extra $(-1)^{\D}$
originate from the ambiguity due to  trivialization
of $det\;ind(\Fs{\eufm{D}}^x)$   in the  relative orientation
between the instanton moduli space and
the (abelian) Seiberg-Witten moduli spaces.

It is easy to include the observable $\hat u$ and compute
the  general correlation function
\eqn\nuu{
\left< \exp(\hat v + \tau \hat u)\right>_{m,\eufm{c},k}.
}
We replace $\hat u = -\Fr{1}{8\pi^2}\tr \phi^2$ with
its value in branch (ii),
$\hat u(ii) = m^2/4\pi^2$.
A similar manipulation leads to
\eqn\cst{\eqalign{
\left< \exp(\hat v + \tau \hat {u})\right>_{m,\eufm{c},k}
=
&- \left(-\Fr{2\pi}{m}\right)^{d_0(\eufm{c},k)}
\biggl(
\left< e^{\hat v + \tau \hat {u}} \right>_{k}
\cr
&
-2^{2+\Fr{1}{4}\left(7\chi + 11\s\right)}
\left(\Fr{2\pi}{m}\right)^{d(k)}
\sum_x n_x
e^{\Fr{m}{2\pi} (v\cdot x) +\Fr{m^2}{4\pi^2}
\left(\Fr{v\cdot v}{2}\right) +\tau\Fr{m^2}{2\pi^2}}
\biggr).
}
}
We get
\eqn\csu{
\left< \exp(\hat v + \tau \hat {u})\right>_{k}
=
2^{2+\Fr{1}{4}\left(7\chi + 11\s\right)}
\sum_{r+s=0}^{[d(k)/2]}
\Fr{\left(\Fr{v\cdot v}{2}\right)^r (2\tau)^{s}}{
(d(k)-2r-2s))!r!s!}
\sum_{x}n_x(v\cdot x)^{d(k)-2r-2s}.
}
In other words,
\eqn\vvb{
\left<\hat v^{d(k)-2s}  \hat u^s \right>_k =
2^{2+\Fr{1}{4}\left(7\chi + 11\s\right)}
\sum_{r=0}^{[d(k)/2 -s]}\Fr{(d-2s)!(2\tau)^s}{(d(k)-2r-2s)! r!}
\left(\Fr{v\cdot v}{2}\right)^r
\sum_{x}n_x(v\cdot x)^{d(k) - 2r -2s}.
}
Note that one can write
$\left< \exp(\hat v + \tau \hat {u})\right>_{k}$
as
\eqn\cssus{
\left< \exp(\hat v + \tau \hat {u})\right>_{k} =
2^{2+\Fr{1}{4}\left(7\chi + 11\s\right)}
\left(\Fr{2\pi}{m}\right)^{d(k)}
\sum_x n_x
e^{\Fr{m}{2\pi} (v\cdot x)
+\Fr{m^2}{4\pi^2}\left(\Fr{v\cdot v}{2}\right) +\tau\Fr{m^2}{2\pi^2}},
}
provided  we take the $zero$th order of $m$ only in
the formal expansion of RHS. Recall that $m$ was assigned
to the $U$-number $2$, so the $U$-number anomaly cancellation
of the path integral of TYM is beautifully summarized in
the above formula.

The correlation function
$\left<\exp({\hat v}+\tau\hat  u) \right>_{\eufm{c},k}$
of the massless TQCD can also be  obtained by collecting
the $zero$th order of $m$ in the formal expansion of
\cst. We have
\eqn\cssus{
\left< \exp(\hat v + \tau \hat {u})\right>_{\eufm{c},k} =
2^{2+\Fr{1}{4}\left(7\chi + 11\s\right)}
(-1)^{d_0(\eufm{c},k)}
\left(\Fr{2\pi}{m}\right)^{d(\eufm{c},k)}
\sum_x n_x
e^{\Fr{m}{2\pi} (v\cdot x) +\Fr{m^2}{4\pi^2}
\left(\Fr{v\cdot v}{2}\right) +\tau\Fr{m^2}{2\pi^2}},
}
provided  we take the $zero$th order of $m$ only.
The above formula summarizes the $U$-number anomaly cancellation
of the path integral of TQCD.
There is a subtlety due to the additional sign factor
$(-1)^{d_0(\eufm{c},k)}$. If we replace  $m$ with $-m$,
we have
\eqn\cssus{
\left< \exp(\hat v + \tau \hat {u})\right>_{\eufm{c},k} =
2^{2+\Fr{1}{4}\left(7\chi + 11\s\right)}
\left(\Fr{2\pi}{m}\right)^{d(\eufm{c},k)}
\sum_x n_x
e^{\Fr{m}{2\pi} (v\cdot x) +\Fr{m^2}{4\pi^2}
\left(\Fr{v\cdot v}{2}\right) +\tau\Fr{m^2}{2\pi^2}}.
}
This shows that the polynomials vanish unless
\eqn\vanishing{
d_0(\eufm{c},k)
\equiv  -k +\Fr{1}{4}(\eufm{c}\cdot\eufm{c}-\s) = 0
\hbox{ mod } 2.
}
Since $\eufm{c}\cdot\eufm{c} = \s \hbox{ mod } 8$,
the polynomial identically vanishes if the  instanton number $k$
is odd.
So the total degree of the polynomial
$\left< \exp(\hat v + \tau \hat {u})\right>_{\eufm{c},k}$
increases as $12 \BZ$ rather
than $6 \BZ$. Note that the degree of Donaldson's polynomial
increases as $8\BZ$.\foot{
We count  the degrees of $\hat v$ and $\hat u$ by $2$
and $4$, repectively.
The dimensions of the moduli
space $\CM(\eufm{c},k)$  is proportional to
the instanton number $k$ by the factor
$2(4-1) = 6$. The dimension of the moduli space $\CM(k)$
is proportional to the instanton number $k$ by the factor
$2\cdot 4 = 8$.
These factors are closely related to the anomaly
free discrete subgroup of the global $U(1)_\CR$ symmetry
of the underlying physical theories. For the $SU(2)$ and
$N_f=0$ theory, the discrete subgroup is $Z_8$.
On the other hand, for $N_f=1$ the anomaly   free
discrete subgroup is $Z_{12}$ rather than $Z_6$ \SWb.
After twisting the $U(1)_\CR$ charge becomes
the $U$-number.
The property that
$\left< \exp(\hat v + \tau \hat {u})\right>_{\eufm{c}, k}$
vanishes for odd $k$ implies the fact that  the anomaly   free
discrete subgroup is $Z_{12}$ rather than $Z_6$ and vice versa.
}
Thus, for an even instanton number $k$, we have
\eqn\csv{
\left< \exp(\hat v + \tau \hat {u})\right>_{\eufm{c},k}
=2^{2+\Fr{1}{4}\left(7\chi + 11\s\right)}
\sum_{r+s=0}^{[d(\eufm{c},k)/2]}
\Fr{\left(\Fr{v\cdot v}{2}\right)^r (2\tau)^{s}}{(d(\eufm{c},k)-2r-2s))!r!s!}
\sum_{x}n_x(v\cdot x)^{d(\eufm{c},k)-2r-2s},
}
while for an odd instanton number $k$, we have
\eqn\csvta{
\left< \exp(\hat v + \tau \hat {u})\right>_{\eufm{c},k} =0.
}

By substituting \csu\ to \cst\
we also get
\eqn\cstab{
\eqalign{
\left< e^{\hat v + \tau \hat {u}}\right>_{m,\eufm{c},k}
=
&2^{2+\Fr{1}{4}\left(7\chi + 11\s\right)}
\left(-\Fr{2\pi}{m}\right)^{d_0(\eufm{c},k)}
\biggl[\left(\Fr{2\pi}{m}\right)^{d(k)}
\sum_x n_x
e^{\Fr{m}{2\pi} (v\cdot x) +\Fr{m^2}{4\pi^2}
\left(\Fr{v\cdot v}{2}\right) +\tau\Fr{m^2}{2\pi^2}}
\cr
&\qquad -
\sum_{r+s=0}^{[d(k)/2]}
\Fr{\left(\Fr{v\cdot v}{2}\right)^r (2\tau)^{s}}{
(d(k)-2r-2s))!r!s!}
\sum_{x}n_x(v\cdot x)^{d(k)-2r-2s}\biggr].
}
}

The final step of our computation is
to construct the generating functional
\eqn\genea{
\left<\exp\left( \hat v + \l \hat u\right)\right>
= \sum_k \left< \exp(\hat v + \l \hat {u})\right>_{k}.
}
Note that
\eqn\vvb{
d(k) = 4k -3\D,\qquad
n_{-x} = (-1)^\D n_x.
}
The result is
\eqn\goala{\eqalign{
\left<\exp\left( \hat v + \tau \hat u\right)\right>
=
& 2^{1+\Fr{1}{4}\left(7\chi + 11\s\right)}
   \biggl[\exp\left(\Fr{v\cdot v}{2} + 2\tau\right)\sum_x n_x e^{v\cdot x}
\cr
& \phantom{........}+ i^{\D}\exp
   \left(-\Fr{v\cdot v}{2} - 2\tau\right)\sum_x n_x e^{-iv\cdot x}
   \biggr],
\cr
}}
which is the formula \goal\ of Witten.
Note that
\eqn\simplea{
\left(\Fr{d^2}{d\tau^2} -2^2\right)
\left<\exp\left( \hat v + \tau \hat u\right)\right>
=0,
}
or equivalently
\eqn\equia{
\left<\left(\hat u^2-2^2\right)\hat z\right>=0,\qquad\hbox{for any } \hat z.
}
The above two equivalent conditions are called the simple
type conditions.
One can define the Donaldson series $\msbm{D}$
\eqn\aaf{\eqalign{
\msbm{D}(v)
&\equiv\Fr{1}{2} \left(1 + \Fr{1}{2}\Fr{d}{d\tau}\right)
\left<\exp(\hat v + \tau \hat u)\right>\biggl|_{\tau =0}
\equiv \Fr{1}{2}\left<(1 + \Fr{\hat u}{2})\exp(\hat v)\right>
\cr
&= 2^{1+\Fr{1}{4}(7\chi + 11\s)}
\exp\left({\Fr{v\cdot v}{2}}\right)\sum_{x} n_x e^{v\cdot x}.
}
}
Kronheimer and Mrowka proved that\KMa
\eqn\aag{
\msbm{D}(v)=
\exp\left({\Fr{v\cdot v}{2}}\right)\sum_{x} a_x e^{v\cdot x},
}
where $a_x$ is a (non-zero) rational number
and $x$ is a basic class which is an integral lift of the
second Stifel-Whitney class $w_2(X)$ on $X$.
The main predictions of the formula \goal\ of Witten  are that
the basic class of Kronheimer-Mrowka is the Seiberg-Witten class
and that
\eqn\aagb{
a_x = 2^{1 + \Fr{1}{4}(7\chi + 11\s)} n_x.
}
The formula \aaf\  confirms those predictions.\foot{
To complete the proof, the factor $2^{\Fr{1}{4}(7\chi + 11\s)}$
should be derived without referring to the known mathematical
results. A step has been made by Witten in \WittenS.
Our results suggest that the factor is universal for the
family of TQCD with $N_f =0,\ldots, 4$. Then, the factor may be
derived by imposing the exact self-duality of the critical
($N_f=4$) theory \SWb\VW.
}

The generating functional for TQCD is defined by
\eqn\geneb{
\left<\exp\left( \hat v + \l \hat u\right)\right>_\eufm{c}
= \sum_k \left< \exp(\hat v + \tau \hat {u})\right>_{\eufm{c}, k}.
}
Note that
\eqn\vvba{
d(\eufm{c},k) = 3k -3\D +\Fr{1}{4}(\eufm{c}\cdot\eufm{c} -\s),
\qquad n_{-x} = (-1)^\D n_x,
}
and the summation over $k$ in \geneb\ is
only for even $k$.
We have
\eqn\goalb{\eqalign{
 \biggl< \exp\left(\hat{v} + \tau\hat{u}\right)\biggr>_\eufm{c}
=
&\Fr{2}{3}\cdot 2^{1+\Fr{1}{4}(7\chi + 11\s)}
\biggl[
\exp\left(\Fr{v\cdot v}{2} +  2\tau\right)
\sum_{x} n_x e^{(v\cdot x)}
\cr
&\qquad
+ (-1)^\D e^{\Fr{\pi i}{12}(\eufm{c}\cdot\eufm{c}-\s)}
\exp\left(
e^{-\Fr{2\pi i}{3} }\left(\Fr{v\cdot v}{2} +  2\tau\right)
\right)
\sum_{x} n_x e^{e^{-\Fr{\pi i}{3}} (v\cdot x)}
\cr
&\qquad
+ e^{\Fr{\pi i}{6}(\eufm{c}\cdot\eufm{c}-\s)}
\exp\left(
e^{-\Fr{4\pi i}{3}}\left(\Fr{v\cdot v}{2} +  2\tau \right)
\right)
\sum_{x} n_x e^{e^{-\Fr{2\pi i}{3}} (v\cdot x)}
\biggr].
}
}
Note that
\eqn\simpleb{
\left(\Fr{d^3}{d\tau^3} -2^3\right)
\left<\exp\left( \hat v + \tau \hat u\right)\right>_\eufm{c}
=0,\qquad
\left<\left(\hat u^3-2^3\right)\hat z\right>_\eufm{c}=0,
\qquad\hbox{for any } \hat z.
}
This is a generalized simple type condition for the polynomial
invariants \goalb.
Following Kronheimer and Mrowka,
one can define the Donaldson series $\msbm{D}(v)_\eufm{c}$
as
\eqn\aafb{
\msbm{D}(v)_\eufm{c}
\equiv \Fr{1}{2} \left(1 + \Fr{1}{2}\Fr{d}{d\tau}
+ \Fr{1}{2^2}\Fr{d^2}{d\tau^2}\right)
\left<\exp(\hat v + \tau \hat u)\right>_\eufm{c}\biggl|_{\tau =0}
\equiv
\Fr{1}{2}\left<\left(1 + \Fr{\hat u}{2} +\Fr{\hat u^2}{2^2}\right)
\exp(\hat v)\right>_\eufm{c}.
}
We have
\eqn\dosse{
\msbm{D}(v)_\eufm{c}
= 2^{1+\Fr{1}{4}(7\chi + 11\s)}
\exp\left({\Fr{v\cdot v}{2}}\right)\sum_{x} n_x e^{v\cdot
x}.
}
Thus we get the same Donaldson series.

\newsec{Relations with the Physical Theory}

In  paper \SWa, Seiberg and Witten studied the exact low energy
effective theory of the $N=2$ supersymmetric $SU(2)$ Yang-Mills
theory on the flat $4$-manifolds.
It turns out that the exact low energy effective theory
can be determined by an analytic pre-potential
which can be expressed in terms of an auxiliary elliptic
curve varying over the quantum moduli space
which parametrizes the different vacua.
The elliptic curve is given by
\eqn\curve{
y^2 = (x^2 -\L^4)(x -\hat u),
}
where $\L$ is the dynamically generated scale of the theory.
In the finite region of the quantum moduli space,
there are  two singularities at $\hat u = \pm \L^2$.
At these two singular points, a new massless particle appears,
which forms $N=2$ supersymmetric $U(1)$ hypermultiplets.
Furthermore, the theory is weakly coupled near the
singularities.

The above two singular points also correspond to
the two vacua of the $N=1$ theory.
For a K\"{a}hler manifold with $b_2^+ > 1$, one can perturb
the theory by adding $N=2$ breaking, but $N=1$ preserving
mass term related to the holomorphic two-form \WittenB.
In the large scale limit of the K\"{a}hler metric, the
dominant contributions to the path integral
only come from these two points.
Witten argues that a similar thing  is happening in
a simple type manifold and that the contributions
only come from a neighborhood of singular points \WittenC.
In computing the path integrals of the TYM theory,
one expands all the operators in terms of operators in
the low energy effective theory.
The simple type condition \equia\ arises when one replaces
the operator $\hat u$ in terms of the $c$-number
$\pm \L^2$.

In this paper, we computed the topological correlation
functions of the twisted $N=2$ supersymmetric Yang-Mills
theory coupled with hypermultiplet having the bare mass.
Our purely semi-classical computation shows that
the path integral of  the TYM theory (Donaldson invariants)
is expressed in terms of the branch (ii) contributions
due to the $U(1)$ massless hypermultiplet.
Note that such a dramatic localization of the path integral
appears in the large scale limit of the metric.
In that limit, the metric on the Riemann manifolds
$X$ becomes everywhere nearly flat.
Thus, the topologically
equivalent description of the twisted $N=2$ SYM theory
can also be the physically equivalent description of
the untwisted theory in the low energy.
This is  reversing the logic of Witten.
An interesting comparison is that we replace
$\hat{u}$ with the bare mass of the hypermultiplet instead of
the dynamically generated scaling parameter.

Our concrete and direct computation, then, clearly predicts
the vacuum structure of the underlying physical theories.
The simple type conditions \equia\ and \simpleb\
can be viewed as the predictions of the singularities
in the quantum moduli spaces.
The vacuum structure of the underlying physical theory
of our model was also determined  by Seiberg and Witten  \SWb.
The simple type condition \simpleb\ is identical to the
locus of the $3$ singularities in the quantum moduli space
of the $N=2$ $SU(2)$ SYM theory coupled with
one hypermultiplet. Our results can be viewed as
a non-trivial check of their solutions.
Our result also suggests that there should be some intriguing
structure hidden in the solutions of Seiberg-Witten.
Note also that  the TYM theory and TQCD are governed
by the same data and define the same Donaldson series.
They belong to the same universality class.
We believe those strange interrelations are originated
from the critical theory $N_f = 4$.

The results in this paper can be generalized to the general
gauge group. In the second paper of the series \HPPb,
we will completely determine the topological correlation
functions of $SU(N_c)$ TQCD on a simple type manifold
coupled with hypermultiplets in the
fundamental representation having the bare mass.
Similarly to this paper, we can also obtain the invariants
of the massless theory and the $SU(N_c)$
Donaldson-Witten invariants as well.
The relation between the vacuum structures of underlying
physical theories and the generalized version of the
simple type conditions for those invariants,
as well as the universality of the
Donaldson series all remain unchanged.

We conclude this paper with  an interesting question.
Fintushel and Stern determined the blowup formula for the
$SU(2)$ Donaldson invariants \FS. Surprisingly, their
formula is stated in terms of the same elliptic curve \curve\
which parametrizes the quantum moduli space.
They also pointed out the relation between the
simple type condition \equia\ and the discriminant locus
of the curve.  An analogous blowup formula
might be constructed for the invariants \goalb.
It will be interesting to see if such a formula
recovers the elliptic curve for $N_f=1$ theory \SWb.

After this work was completed,
the announced paper \PTb\ of  Pidstrigach and Tyurin
and   a related work \LMb\ of Labastida and Mari\~{n}o with this paper
appeared. In \LMb , Labastida and Mari\~{n}o calculated the topological
correlation functions of the $SU(2)$ theory with $N_{f}=1$ using the
physical method on a spin manifold with the canonical  $spin_c$ structure.
They also informed us that the twisting of N=2 hypermultiplets were
considered in \Lc\ and
further
elaborated in \Ld\ .

\ack

We are grateful to  Robert Dijkgraaf,  Timothy Hollowood,
and Herman Verlinde  for useful discussions and suggestions.
We are also grateful  to Jorn Verwey for proof-reading on this
manuscript. SH and JSP would like to thank the organizers
of Jerusalem Winter school,   ICTP and the
organizers of spring school for hospitality,
where  part of this work was done
and the main results  have been announced
{}.
The work of SH  is supported in part by
the Basic Science Research
Institute Program,
Ministry of Education Project No. BSRI-95-2425.
The work of JP  is supported in part by
U.S. Department  of Energy Grant No. DE-FG03-92-ER40701.
JSP is grateful to the Isaac Newton Institute, the
University of Amsterdam  and Caltech for hospitality.

\vfill\eject

\appendix{A}{}

In this appendix, we briefly explain the relation
between  the $2m$-point Green's function
$G(x_1,\ldots,x_{2m})$ for a free field theory  and
a multilinear form $Q^{(m)}$ on $H_2(X)$.
We will calculate the path integral of  the simplest possible
quantum field theory.
The calculations in this paper are just some slightly
elaborated variations of the following model.
We will not consider the supersymmetric case.

Let $X$ be a compact oriented Riemann four-manifold.
The intersection form $Q$ is a bilinear form
\eqn\appa{
Q: H_2(X;\BZ)\times H_2(X;\BZ) \rightarrow \BZ.
}
Let $\S_1, \S_2 \in H_2(X;\BZ)$ be closed surfaces representing
two dimensional homology cycles. The intersection number
is defined by the algebraic sum of the number of transverse
intersection points counted with a sign $\pm$ depending
on the orientation near an intersection point $p$
\eqn\appb{
TX_p = T_p\S_1\oplus T_p\S_2.
}
We used notation $Q(\S_1,\S_2) = v_1\cdot v_2$. This definition
also makes sense for the self-intersection number $v\cdot v$;
one can perturb two closed surfaces within
the same homology class  such that they intersect transversely
at points.
The multi-linear form $Q^{(m)}$ on $H_2(X)$ defined by \DK\
\eqn\appc{
Q^{(m)}(\S_1,...,\S_{2m})
=\Fr{1}{2^m m!}\sum_{\s\in S_{2m}}
Q(\S_{\s(1)},\S_{\s(2)})\times\ldots\times
Q(\S_{\s(2m-1)},\S_{\s(2m)}),
}
where $Q(.,.)$ denotes the intersection form.
For example
\eqn\appd{
Q^{(2)}(\S_1,\ldots,\S_4)
= (\S_1\cdot \S_2)(\S_3\cdot\S_4) + (\S_1\cdot \S_3)(\S_2\cdot\S_4)
 + (\S_1\cdot \S_4)(\S_2\cdot\S_3).
}

Now we consider a simple Gaussian integral
\eqn\appe{
Z[J] =\int \CD \phi e^{-\Fr{1}{2}
\int(\phi(x), A \phi(x)) + \int(J(x) ,\phi(x))}.
}
The Gaussian integral over $\phi$ gives
\eqn\appf{
Z[J]= [det(A)]^{-1/2} e^{\Fr{1}{2}\int(J, A^{-1} J)}
= [det(A)]^{-1/2} e^{\Fr{1}{2}\int d^4\!x d^4\!y  J(x)\D_F(x-y) J(y)},
}
where $\D_F(x-y)$ is called the Feynman propagator, defined
by
\eqn\appg{
\D_F(x-y) = \int\Fr{d^4\!p}{(2\pi)^4}
\Fr{e^{-i p\cdot(x-y)}}{ {\widetilde{( A^{-1})}}},
}
where $p$ is the Fourier transformation variable (four-momentum)
for $x$ and $\tilde{A}$ is the Fourier transform of $A$.
One usually defines
\eqn\apph{
\left< \phi(x_1)\ldots \phi(x_n)\right>
= \Fr{\d^{n}}{\d J(x_1)\cdots \d J(x_n)} Z(J)|_{J=0}
=G(x_1,\ldots,x_n).
}
Now we expand $Z[J]$
\eqn\appi{
Z[J] = [det(A)]^{-1/2} \sum_{n=0}^{\infty}
\Fr{1}{n!} \int d^4\!x_1\cdots d^4\!x_n
G_n(x_1,\ldots,x_n) J(x_1)\ldots J(x_n).
}
By inspection, we find the odd order
Green's functions vanish
and
\eqn\appj{
G_2(x_1,x_2) = \D_F(x_1-x_2).
}
One can also find that
\eqn\appk{
G_{2n}(x_1,...,x_{2m})
=\Fr{1}{2^m m!}\sum_{\s\in S_{2m}}
G_2(x_{\s(1)},x_{\s(2)})\times\ldots\times
G_2(x_{\s(2m-1)},x_{\s(2m)}).
}

Now we assume that the operator $A$ is the identity such that
$G_2(x,y) = \d^{(4)}(x,y)$.
Now we assume that $J(x)$ is a de Rham current supported
on a closed surface $\S$ representing a homology cycle
such that $\int_X(J,\phi) =\int_\S\hat{j}(\phi)$.
We can write
\eqn\appl{
Z[J] = \left< e^{\int_\S \hat\phi(x)}\right>
=  e^{\Fr{1}{2}\int d^4\!x d^4\!y  J(x)\d^{(4)}(x-y)J(y)}.
}
Near an intersection point $p$,
we can write the exponent
\eqn\appm{
\int d^4\!x d^4\!y  J(x)\d^{(4)}(x,y) J(y)
=\int_{\S_1\times\S_2} d^2\!x d^2\!y \d^{(4)}(x,y)
= \pm 1,
}
where  $\pm 1$ is determined by the orientation \appb.
Thus we have
\eqn\appn{
Z[J] = e^{\Fr{v\cdot v}{2}}.
}
We can also use the version \appi, we have
\eqn\appm{
Z[J] =  \sum_{n=0}^{\infty}
\Fr{1}{(2n)!}Q^{(n)}(\S_1,...,\S_{2n})
=  \sum_{n=0}^{\infty}
\left(\Fr{1}{2^n n!}q^n\right),
}
where $q = v\cdot v$.
When the surfaces $\S_i$ belong
to the different homology classes, the notation
$Q^{(n)}(\S_1,\ldots,\S_{2n})$ will be more appropriate.

\listrefs
\bye